\documentclass[fleqn,usenatbib]{mn2e}
\usepackage{times}
\usepackage{amsmath}
\usepackage{graphicx}
\usepackage{booktabs} 
\usepackage{dcolumn}
\usepackage{fixltx2e} 

\def\arcsec{$^{\prime\prime}$}

\def\vecp{{\bmath p}}
\def\vecf{{\bmath f}}
\def\vecbeta{{\bmath \beta}}
\def\vecphi{{\bmath \phi}}

\def\mr{$\epsilon_{\rm sys}$}
\def\mar{$\epsilon_{\rm mar}$}
\def\semr{${\rm SE}(\epsilon_{\rm sys})$}
\def\CI{$|{\rm CI}|$}
\def\Con{Con}

\def\logg{$\log g$}
\def\feh{[Fe/H]}
\def\a0{A_0}
\def\av{A_{\rm V}}
\def\mv{M_{\rm V}}

\def\r0{R_0}

\def\Msol{M$_{\odot}$}

\def\ilium{{\sc ilium}}

\title [Bayesian stellar parameter inference]{Bayesian inference of stellar parameters and interstellar extinction using parallaxes and multiband photometry}

\author[C.A.L.\ Bailer-Jones]
{C.A.L.\ Bailer-Jones\thanks{Email: calj@mpia.de}\\
Max-Planck-Institut f\"ur Astronomie, K\"onigstuhl 17, 69117 Heidelberg, Germany\\
Laboratory for Particle Physics and Cosmology, Department of Physics, Harvard University, 38 Oxford Street, Cambridge, MA 02138, USA}

\begin{document}

\date{{\em Submitted 21 July 2010; Revised version submitted 29 August 2010; Accepted 10 September 2010}}

\maketitle

\label{firstpage}

\begin{abstract} 
Astrometric surveys provide the opportunity to measure the absolute magnitudes of large numbers of stars, but only if the individual line-of-sight extinctions are known. Unfortunately, extinction is highly degenerate with stellar effective temperature when estimated from broad band optical/infrared photometry. To address this problem, I introduce a Bayesian method for estimating the intrinsic parameters of a star and its line-of-sight extinction. It uses both photometry and parallaxes in a self-consistent manner in order to provide a non-parametric posterior probability distribution over the parameters.  The method makes explicit use of domain knowledge by employing the Hertzsprung--Russell Diagram (HRD) to constrain solutions and to ensure that they respect stellar physics.  I first demonstrate this method by using it to estimate effective temperature and extinction from $BVJHK$ data for a set of artificially reddened Hipparcos stars, for which accurate effective temperatures have been estimated from high resolution spectroscopy.  Using just the four colours, we see the expected strong degeneracy (positive correlation) between the temperature and extinction. Introducing the parallax, apparent magnitude and the HRD reduces this degeneracy and improves both the precision (reduces the error bars) and the accuracy of the parameter estimates, the latter by about 35\%.  The resulting accuracy is about 200\,K in temperature and 0.2\,mag in extinction.  I then apply the method to estimate these parameters and absolute magnitudes for some 47\,000 F,G,K Hipparcos stars which have been cross-matched with 2MASS.  The method can easily be extended to incorporate the estimation of other parameters, in particular metallicity and surface gravity, making it particularly suitable for the analysis of the $10^9$ stars from Gaia. 
\end{abstract}


\section{Introduction \label{sect:introduction}}

The upcoming astrometric survey Gaia will provide accurate parallaxes (better than 10\%) for about one hundred million stars out to 10\,kpc, representing an enormous increase in the number of stars for which accurate distances can be derived (e.g.\ Mignard \& Drimmel~\citealp{aoresponse}, Lindegren et al.~\citealp{lindegren08}, Bailer-Jones~\citealp{cbj09}).
When combined with a measurement of the apparent magnitude, $m$, this allows us to estimate the
absolute magnitude, $M$, of a star via the simple geometric relationship 
\begin{equation}
m + 5\log\varpi \,=\, M + A - 5
\label{eqn:ma_constraint}
\end{equation}
where $\varpi$ is the parallax (in arcseconds) and $A$ is the interstellar extinction in magnitudes.  Fundamental stellar parameters are usually inferred using just the spectral energy distribution (SED), yet obviously the parallaxes provide an important additional constraint on $M$. 
Yet the above equation can only be applied in a simple manner to deduce $M$ if $A$ is known. In principle it too may be estimated from the SED, but in practice there is a strong degeneracy between extinction and effective temperature ($T$) in optical/near-infrared multiband photometry which limits the accuracy with which either can be estimated. Such a degeneracy exists with the very low resolution Gaia spectroscopy (Bailer-Jones~\citealp{cbj10a}).  Moreover, $M$ is a strong function of $T$, so we cannot estimate $A$ and $T$ from the spectrum and then expect equation~\ref{eqn:ma_constraint} to give a consistent solution for $M$.

A common way to avoid this dilemma is to assume a value for $A$ -- often zero or a value from an extinction map -- but this is rarely valid and certainly not admissible for a deep, all sky survey such as Gaia.  The basic issue is that $A$ and $T$ are not independent of parallax and apparent magnitude. The solution is to solve for $A$ and $T$ simultaneously using both the SED and the parallax/apparent magnitude. This must be done probabilistically in order to properly characterize the intrinsic degeneracy between $A$ and $T$.

This paper introduces a general way to do this which uses Bayes theorem to ensure that all of the information is taken into account self-consistently. The basic idea is to estimate $P(A,T | \vecp, q)$, the posterior probability density function (PDF) over the two parameters given two pieces of information, the normalized SED, $\vecp$, and the quantity $q = m + 5\log\varpi$.  This normalized SED describes just the shape of the SED, ignoring the overall flux level. It is used in a conventional, multivariate, forward modelling approach to compare the data with a set of labelled templates in order to obtain $P(\vecp | A,T)$, the likelihood over $A,T$.  The quantity $q$ constrains the sum $M+A$ (equation~\ref{eqn:ma_constraint}). Used alone it can do little other than place plausible, but not very useful, limits on extreme values of $A$ and $M$.  I therefore explicitly incorporate the knowledge embodied in the Hertzsprung--Russell Diagram (HRD), the distribution of stars in the $(M,T)$ plane.  The physics of stellar structure forbids stars from occupying large areas of this plane, and the nature of stars' structure and changing rates of evolution mean that the remaining parts are far from being uniformly populated.  This well-established information should not be ignored when inferring astrophysical parameters (APs). The method makes uses of this information in a consistent and quantitative probabilistic framework.

In section~\ref{sect:method} I describe the method in detail and derive the basic equations.  I demonstrate it in section \ref{sect:fitvald} by using it to estimate parameters for a set of 5280 stars covering a range of $A$ and $T$ using $BVJHK$ photometry and parallax. These data are based on 880 Hipparcos stars (ESA \citealp{hipcat}) for which effective temperatures were estimated by Valenti \& Fischer~\citep{vf05} from echelle spectra. I artificially redden the data in order to introduce extinction variance.
As the ``true'' parameters of these data are known, it can be shown that the method improves the parameter estimation accuracy compared to using just the four colours. I then apply this method in section~\ref{sect:hip2mass} to ``blindly'' estimate $A$ and $T$ for 85\,000 Hipparcos stars.

The motivation for this work is to make best use of the parallax in order to improve the estimation of stellar astrophysical parameters. In principle one could add $q$ as another input alongside $\vecp$ to a pattern recognition algorithm such as a neural network or a support vector machine. But such tools fail to recognize the astrophysical significance of this extra input, and unpublished tests by the Gaia group at MPIA show that this approach indeed does not work.

The present paper is not the first to combine astrometric and SED data for stellar parameter estimation in a probabilistic manner. But it is, to the author's knowledge, the first to introduce extinction as a free parameter and to include the HRD in the estimation process.  Many authors first derive $T$ and then use the parallax to derive $M$ assuming zero extinction.
Alternatively $T$ is derived assuming a value for extinction, a prerequisite for many methods.
For example, Takeda et al.~\citep{takeda07} use the inferred stellar parameters ($T$, \logg, and \feh) from Valenti \& Fischer~\citep{vf05} to predict the parallax, and then use this in a likelihood model together with evolutionary tracks to infer luminosity, mass and age. This approach does not use the HRD prior nor does it solve for extinction (although assuming zero extinction is probably a valid assumption for these very nearby stars).  Pont \& Eyer~\citep{pont04} and Joergensen \& Lindegren~\citep{jorgensen05} develop Bayesian methods for estimating stellar ages, but they both assume that $T$ is already known and that $A$ is either known or zero. However, this overlooks the fact that in most large surveys $T$ must be estimated from multiband photometry or low resolution spectroscopy, and that it is degenerate with $A$ (which is rarely known independently).  Estimating both $T$ and $A$ is non-trivial so they should not be considered as ``input data'' for an inference. Rather they should be part of that inference in order that their uncertainties and degeneracies be correctly propagated.

Using the Bayesian framework we can also turn the problem around in order to estimate, for example, stellar distances given some measured properties of the stars. Burnett \& Binney~\citep{burnett10} recently outlined a method for obtaining ``spectroscopic parallaxes'' in this way.

\section{Theory}\label{sect:method}

\subsection{Problem statement}\label{sect:statement}

We would like to determine the probability density function over the stellar parameters, $\vecphi$, given measurements of the spectral energy distribution, apparent magnitude and parallax.  I will restrict the problem to $\vecphi = (\a0, T)$, i.e.\ to determining $P(\a0, T | \vecp, q)$, although a generalization is straight forward (see section~\ref{sect:extensions}). Before deriving an expression for this in section~\ref{sect:theory}, I must first introduce and explain a few concepts.  The method involves calculating likelihoods based on forward modelling of the SED (sections~\ref{sect:fm} and~\ref{sect:likelihood}) for which we need a template grid which shows variance in effective temperature and extinction.  The parallax and apparent magnitude are then introduced using the $q$ constraint and the HRD prior (sections~\ref{sect:qconstraint} and~\ref{sect:hrdprior}).  Table~\ref{notation} summarizes the main notation I use.

\begin{table}
\begin{center}
\caption{Notation \label{notation}}
\begin{tabular}{ll}
\hline
$V$  & apparent magnitude in the $V$ band (mag)\\
$\mv$ & absolute magnitude in the $V$ band (mag)\\
$\av$ & extinction in the $V$ band (mag) \\
$\a0$ & extinction parameter (mag) \\
$\r0$ & selective extinction parameter \\
$T$  & stellar effective temperature (K) \\
$Z$  & stellar metallicity (fraction) \\
$\varpi$ & parallax (arcsec) \\
$q$  & $\equiv V + 5\log\varpi$ (mag) \\
$\vecp$ & normalized spectral energy distribution with elements $\{p_i\}$ \\
$\vecphi$ & set of stellar astrophysical parameters (APs) \\
$P$ & probability density \\
$\log$ & base 10 logarithm\\
\hline
\end{tabular}
\end{center}
\end{table}

\subsection{Interstellar extinction}\label{sect:extinction}

In order to construct a grid showing a range of interstellar extinction, we need to adopt an extinction law. I adopt the widely-used form
from Cardelli et al.\ \citep{cardelli89}.
This gives the
monochromatic extinction in a narrow band at wavelength $\lambda$ in terms of two extinction parameters $\a0$ and $\r0$ as
\begin{equation}
A_{\lambda} \,=\, \a0 [a_{\lambda} + b_{\lambda}/\r0] \ ,
\label{eqn:extlaw}
\end{equation}
where $a_{\lambda}$ and $b_{\lambda}$ are fixed polynomials. $\a0$ is frequently written as $\av$ in this equation, but this is confusing because
$\a0$ {\em is not the extinction in the $V$ band}. The extinction in the $V$ filter (or indeed, any filter) with pass band function $h_{\lambda}$ is a consequence of integration over the stellar spectral energy distribution, $F_{\lambda}$, i.e.\
\begin{equation}
\av \,=\, -2.5 \log \left( \frac{ \int F_{\lambda} h_{\lambda} 10^{-0.4A_{\lambda}} d \lambda }{ \int F_{\lambda} h_{\lambda} d \lambda }  \right) \ .
\label{eqn:extint}
\end{equation}
Thus $\av$ depends on the spectral energy distribution of the specific star observed, and hence on its intrinsic parameters (in particular effective temperature). Two stars with different $T$ will generally have different $\av$ for the same $\a0$.  $\a0$, in contrast, is a property of the interstellar medium only and so is a better parameter with which to characterize the interstellar extinction.

As the $q$ constraint depends fundamentally on $\av$ rather than $\a0$, we need to express the former in terms of the latter.
This will be done in section~\ref{sect:qconstmodel}. It turns out that for F,G,K stars with extinctions up to 3.5\,mag, the difference between $\a0$ and $\av$ is less than 0.2\,mag.

The artificial reddening will be done using using the specific extinction curves from Fitzpatrick~\citep{fitzpatrick99} with $\r0$\,=\,3.1.

\subsection{Forward model}\label{sect:fm}

The forward model predicts the observed stellar spectral energy distribution, $\vecp$, given the stellar astrophysical parameters, $\vecphi$.  How many astrophysical parameters we need to consider for an accurate prediction depends in particular on the type of stars we want to model and on the resolution of $\vecp$.  Note that $\vecp$ is a {\em normalized} SED, i.e.\ it contains no apparent magnitude information.  Here the SED is a set of colours derived from broad band photometry, so I limit the parameters to $\a0$ and $T$, $\hat{\vecp} = \vecf(\a0, T)$. All other APs are assumed either to be fixed ($\r0$) or to have negligible impact on the normalized SED (\feh\ and \logg).  Although \feh\ has a significant and usable effect on  broad $U$-band photometry (e.g.\ Ivezi\'c et al.~\citealp{ivezic08}), its impact on the redder bands considered here is minimal and is neglected. \logg\ is an even weaker parameter (Bailer-Jones~\citealp{cbj10a}) so its variance too is neglected. The method can nonetheless be generalized to incorporate these extra parameters as appropriate.

The forward model is calculated by a smooth fit to a set of templates using the method developed for the \ilium\ algorithm 
(Bailer-Jones et al.\ \citealp{cbj10a}). It involves fitting a two-dimensional smoothing spline (a thin-plate spline) as a function of $\av$ and $T$ for each element of $\vecp$ separately.

\subsection{The likelihood model}\label{sect:likelihood}

The likelihood of the spectral data given the astrophysical parameters is $P(\vecp | \vecphi) = P(\vecp | \a0, T)$. Assuming Gaussian errors on a measurement of $\vecp = (p_1, \ldots, p_i, \ldots, p_I)$ with covariance matrix ${\mathbfss C}_p$, the likelihood model is an $I$-dimensional Gaussian
\begin{equation}
P(\vecp | \vecphi) \, \propto \, e^{-D^2/2}  \,=\,\exp \left(  -\frac{1}{2}[\vecp - \vecf(\vecphi)]^T {\mathbfss C}_p^{-1}[\vecp - \vecf(\vecphi)]  \right)  \ .
\label{eqn:likelihood}
\end{equation}
If the elements of $\vecp$ were uncorrelated then ${\mathbfss C}_p = {\rm diag}(\sigma^2_{p_i})$, where $\sigma_{p_i}$ is the expected error in $p_i$,  so the exponent could be simplified to
\begin{equation}
D^2 = \sum_{i=1}^{i=I} \left[ \frac{p_i - f_i(\vecphi)}{\sigma_{p_i}}  \right]^2 \ .
\end{equation}

\subsection{Parallax/magnitude ($q$) constraint}\label{sect:qconstraint}

As outlined in the introduction, simple geometry and the definition of absolute magnitude and extinction places the following constraint on noise-free quantities
\begin{equation}
V + 5\log\varpi \,=\, \mv + \av - 5 \
\label{eqn:ma_constraint2}
\end{equation}
(I assume we measure the apparent magnitude in the $V$ band, although any other band would do).
The goal is to use this equation to constrain $\mv$ and $\av$
from noisy measurements of parallax and magnitude. To do this we need a noise model.
For brevity define
\begin{equation}
q \, \equiv \, V + 5\log\varpi \ .
\label{eqn:q}
\end{equation}
Since equation~\ref{eqn:ma_constraint2} only holds in the absence of noise, consider the random variable
\begin{equation}
x \, = \, q - (\mv + \av - 5) \ \ .
\label{eqn:x}
\end{equation}
The noise model for $x$ is $P(x | \mv, \av)$, which has expectation value zero and variance
$\sigma_q^2$, the variance in $q$ ($\mv$ and $\av$ are not measured so contribute no noise).  
For simplicity I choose to model this as a one-dimensional Gaussian in $x$, ${\mathcal N}_x(0, \sigma_q)$.  For a given star (fixed $\mv$ and $\av$), $P(x | \mv, \av)$ has its maximum when the measurement $q$ equals $\mv + \av - 5$ (i.e.\ $x=0$). The further a measurement of $q$ is away from this value the less probable it is.  As $q$ is the only measured term in equation~\ref{eqn:x} it follows that $P(x | \mv,\av) = P(q | \mv,\av)$.

\begin{figure}
\begin{center}
\includegraphics[scale=0.48, angle=0]{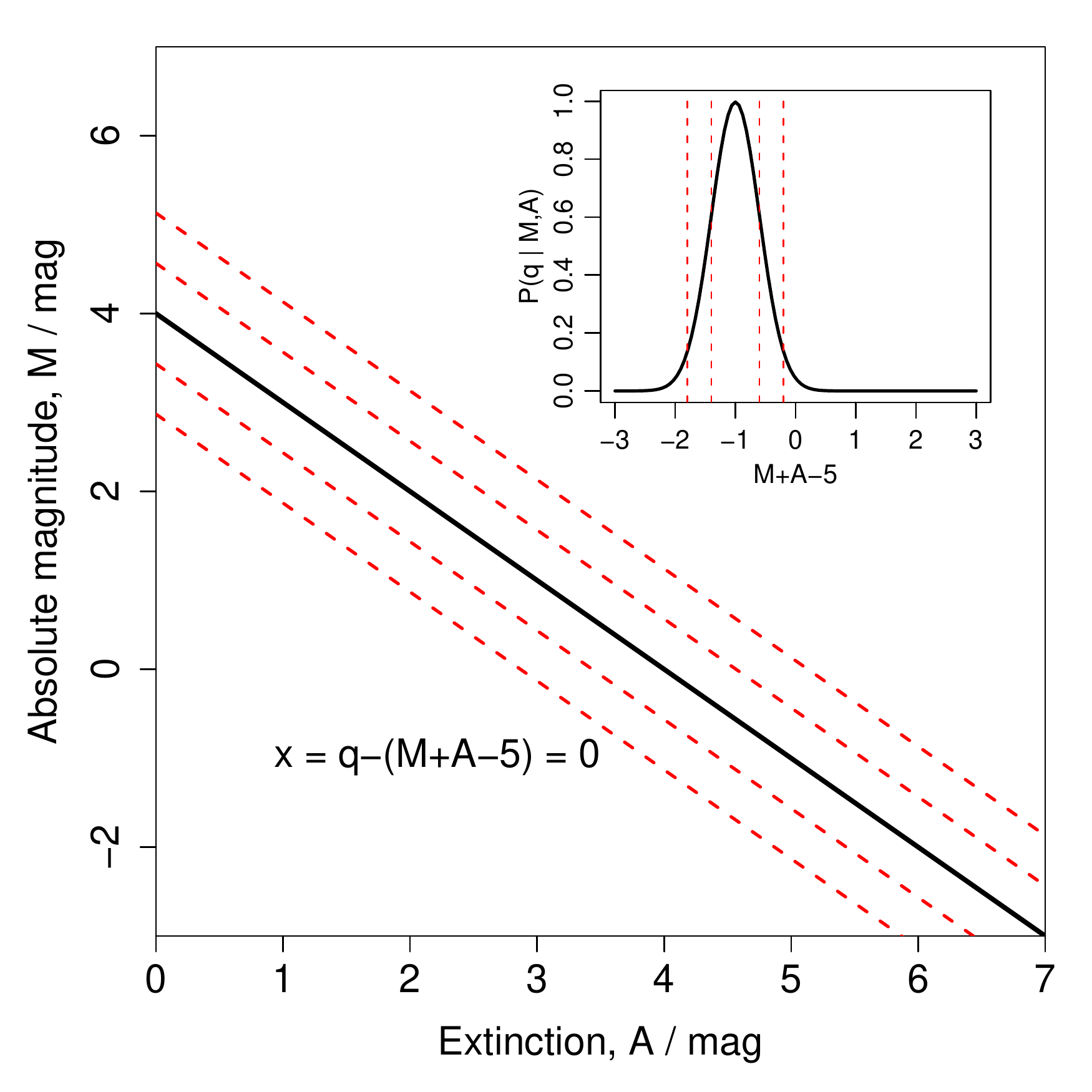}
\caption{Illustration of using the parallax and apparent magnitude ($q=V + 5\log\varpi$) to constrain extinction and absolute magnitude. Here we measure $q=-1$ (which corresponds to a $V=14$ star at 1\,kpc, for example, or to a $V=19$ star at 10\,kpc, etc.). If this were a noise-free measurement, it would constrain the solution $(\mv,\av)$ to lie on the solid black line. But as $q$ is a noisy measurement -- here a Gaussian with $\sigma_q=0.4$ (inset) -- all solutions have a finite probability, decreasing with distance from the line. Specifically, any slice
perpendicular to the line has the Gaussian profile show in the inset panel, the red dotted lines
in both plots showing the 1 and 2 sigma levels for this value of $\sigma_q$.
\label{fig:ma_constraint}}
\end{center}
\end{figure}
 
Now consider $P(q | \mv, \av)$ as a function of $\mv$ and $\av$ for a given measurement $q$, as shown in Fig.~\ref{fig:ma_constraint}. We can think of proposing trial solutions for $\mv$ and $\av$: the further they lie from the solid line, the lower $P(q | \mv, \av)$ (inset in Fig.~\ref{fig:ma_constraint}). How quickly the probability drops off depends on $\sigma_q$. 
With the Gaussian approximation of the noise model for $q$ we have
\begin{eqnarray}
P(q | \mv, \av) \,&=&\, {\mathcal N}_x[0,\sigma_q(V,\varpi)] \\ \nonumber
                          \,&=&\, {\mathcal N}_x[q - (\mv + \av - 5), \sigma_q(V,\varpi)] \ .
\label{eqn:pdf_q}
\end{eqnarray}
This gives a probabilistic constraint on $\mv$ and $\av$ from a measurement of $q$, quantified by the known statistics of the noise in the photometry and parallaxes.  As noted in section~\ref{sect:extinction}, we can write $\av$ as a function of $\a0$ and $T$, so this {\em $q$ constraint} can be written $P(q | \mv, \a0, T)$.

Note that this does not constrain $\mv$ or $\av$ to have astrophysically ``sensible'' values (e.g.\ the line continues to negative $\av$ in Fig.~\ref{fig:ma_constraint}). This may be done by the HRD prior and/or a prior on extinction.

\subsection{Hertzsprung--Russell Diagram (HRD) prior}\label{sect:hrdprior}

The HRD prior, $P(\mv, T)$, gives the relative probabilities of finding stars in different parts of the HRD.  The fact that this $(\mv,T)$ plane is far from being uniformly populated is potentially useful in constraining stellar APs: if $\mv$ were known to lie in some range with some probability, for example, $T$ would correspondingly be constrained.  This is pertinent information independent of the specific photometric or parallax measurement.

The form we adopt for the HRD depends on the assumed stellar population and can be constructed in a number of different ways. We could, for example, take an observed sample and normalize the relative density of stars to give $P(\mv, T)$. Alternatively we could set the probability at each point to be inversely proportional to the speed of evolution of all types of stars through that point. In this article I will construct the HRD prior using a simulated population of stars evolved with a specified star formation rate, initial mass function and metallicity distribution (section~\ref{sect:hrdpriormodel} and Fig.~\ref{hrdmap_vband_comb_z019_8_joinB_forpaper}).

\subsection{Probabilistic combination}\label{sect:theory}

We are now in a position to derive an expression for $P(\a0, T | \vecp, q)$ in terms of quantities we have just introduced.
From Bayes' theorem
\begin{equation}
P(\a0,T | \vecp,q) \,=\, \frac{P(\vecp,q | \a0,T) P(\a0,T)}{P(\vecp,q)} 
\label{eqn:a01}
\end{equation}
and from the rule of joint probabilities
\begin{equation}
P(\vecp,q | \a0,T) = P(\vecp | q,\a0,T) P(q | \a0,T) \ .
\label{eqn:a02}
\end{equation}
As $\vecp$ and $q$ are independent measurements\footnote{Here I only assume that 
$\vecp$ and $q$ are independent when conditioned on $\a0$ and $T$, although normally we would 
further assume them to be unconditionally independent.
This is the case when $\vecp$ is a normalized SED, as then it bears no distance or apparent magnitude information.
}
we can write $P(\vecp | q,\a0,T) = P(\vecp | \a0,T)$.
This and equation~\ref{eqn:a02} allow us to write equation~\ref{eqn:a01} as
\begin{eqnarray}
P(\a0,T | \vecp,q) \,=\, \frac{P(\vecp | \a0,T) \, P(q | \a0,T) \, P(\a0)\, P(T)}{P(\vecp, q)}  
\label{eqn:a04}
\end{eqnarray}
where I have also assumed that $\a0$ and $T$ are unconditionally independent.
The terms $P(\a0)$, $P(T)$, $P(\vecp, q)$ are the inevitable priors over these APs or measurements.
The first term in the numerator is the likelihood (section~\ref{sect:likelihood}). The second term we need to further decompose, plus we want to introduce some dependence on $\mv$ so that we can incorporate the HRD and the $q$ constraint.
A general rule of probability allows us to write this term as a marginalization over $\mv$
\begin{equation}
P(q | \a0,T) \,=\, \int_{M_V} P(q | \mv,\a0,T)\,P(\mv | \a0,T) \, d\mv \ .
\label{eqn:a05}
\end{equation}
The first term in the integral is the $q$ constraint (section~\ref{sect:qconstraint}). As $\a0$ is independent of $\mv$ and $T$, we can rewrite the second term in the integral as
\begin{equation}
P(\mv | \a0,T) \,=\, P(\mv | T)  \,=\, \frac{P(\mv,T)}{P(T)}  \ .
\label{eqn:a06}
\end{equation}
(Another way of thinking about this is to note that given $T$, $\a0$ tells us nothing additional about $\mv$.)
$P(\mv,T)$ is the HRD prior.

Substituting equation~\ref{eqn:a06} into equation~\ref{eqn:a05} and that into equation~\ref{eqn:a04} gives the final result
\begin{align}
\label{eqn:pdfat}
& P(\a0,T | \vecp,q) \,=\, \\ & \underbrace{P(\vecp | \a0,T)}_\text{likelihood} \ \underbrace{\frac{P(\a0)}{P(\vecp, q)}}_\text{priors} \ \ \underbrace{ \int_{\mv} \underbrace{P(q|\mv,\a0,T)}_\text{$q$ constraint}  \underbrace{P(\mv,T)}_\text{HRD prior} d\mv }_\text{HRD/q factor} 
\nonumber
\end{align}
where we see that $P(T)$ has cancelled.  This equation can be seen as a product of three terms. The first term is the likelihood function. The second term comprises priors over the extinction and the data.  Of these, $P(\vecp, q)$ is not relevant (for AP estimation) because the data are already given.  The third term is an integral over two factors: the combined astrometric/photometric noise model ($q$ constraint) and the HRD prior. The integral marginalizes over the unknown $\mv$ leaving a term which is a function of $\a0$ and $T$.

Given measurements of $\vecp$ and $q$ we can sample the terms in equation \ref{eqn:pdfat} on a grid of $\a0$ and $T$ in order to map the full PDF. We can also separately marginalize over $\a0$ and $T$ in order to get one-dimensional PDFs for each AP, i.e.\
\begin{equation}
P(T | \vecp,q) = \int_{\a0} P(\a0,T | \vecp,q) d\a0
\label{eqn:pdfatmarg}
\end{equation}
and likewise for $\a0$. If appropriate we may then summarize this with the mean and a confidence interval.

If we lack information (or don't want to use it) then some terms in equation~\ref{eqn:pdfat} simplify. For example, if we have no measurement of $q$ then we can set the $q$ constraint to a constant. In that case the integral over $\mv$ makes the HRD prior into a prior on just $T$. If we don't want to use an informative prior on the extinction we can set $P(\a0)$ to be constant.  Likewise, if we don't want to use the HRD prior, then this is equivalent to setting $P(\mv,T)$ to a flat distribution (!).  In practice the $q$ constraint is only effective if we use it together with the HRD prior and/or the extinction prior.

Throughout the rest of this paper I will use a uniform extinction prior. As it is separable in equation~\ref{eqn:pdfat}, we can easily imagine the effect of introducing this prior subsequently. I will show two sets of results for $P(\a0,T | \vecp,q)$ based on two different sets of assumptions (priors). The first is a uniform HRD prior and constant $q$ constraint, in which case the posterior PDF is just equal to the likelihood function (renormalized), i.e.\ the APs are inferred using only the spectrum, $\vecp$. I will therefore refer to this as the {\em p-model}. This is the baseline against which I will analyse the effect of using the HRD/q factor, using specific models for the $q$ constraint and HRD prior described in the next section. I will refer to this as the {\em pq-model}. 

It may be useful to recognise that when $\vecp$ and $q$ are unconditionally independent (the normal case), we can interpret equation~\ref{eqn:pdfat} as the combination of two separate estimates of the PDF over $(\a0,T)$ given each of $\vecp$ and $q$. We can see this when we use Bayes' theorem to rewrite the right hand side of equation~\ref{eqn:a04} as
\begin{equation}
P(\a0,T | \vecp,q)\,=\, \frac{P(\a0,T | \vecp) P(\a0,T | q)}{P(\a0,T)} \ .
\label{eqn:a07}
\end{equation}
The p-model is simply $P(\a0,T | \vecp)$.

\section{Model fitting and validation}\label{sect:fitvald}

Let us now use the probabilistic model to estimate the full posterior PDF over effective temperature and extinction for stars with measured five-band photometry and parallaxes.

All calculations are actually done using $\log T$ rather than $T$ -- as then uniform samplings are more appropriate -- but many results will be shown in terms of $T$. (The above theory holds for any monotonic transformation of the variables.) Recall that a small error in $\log T$ of $\delta(\log T)$ corresponds to a fractional error ($\delta T / T$) of about $2.3\delta(\log T)$ .

\subsection{Construction of the labelled data set (``extended catalogue'')}\label{sect:datacon}

\begin{figure}
\begin{center}
\includegraphics[width=0.36\textwidth]{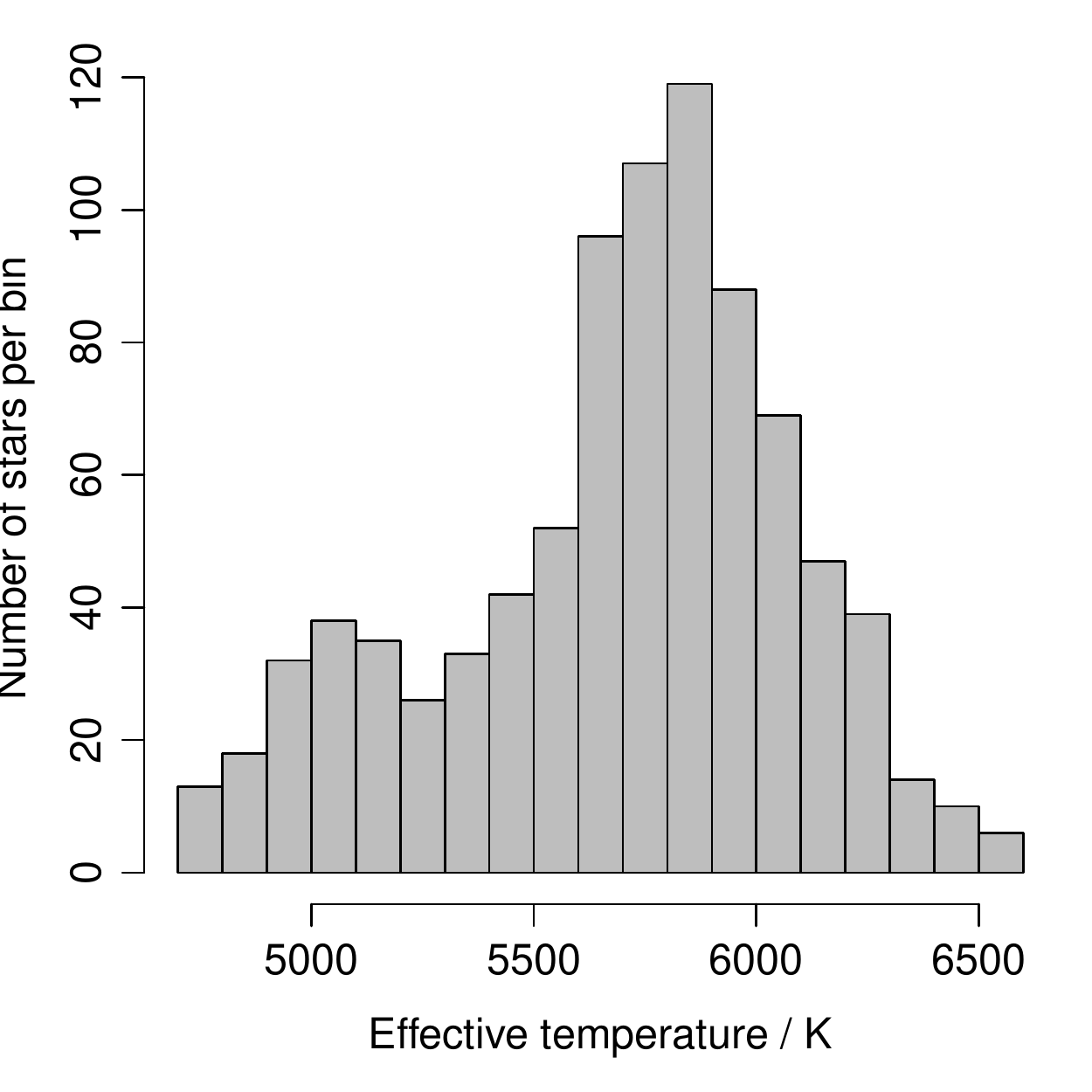}
\caption[]{Distribution of effective temperature of the 880 stars from Valenti \& Fischer \citep{vf05} used to fit and test the model.
\label{v2dat_teff_histogram}}
\end{center}
\end{figure}

In order to validate the performance of the model and to assess whether the HRD/q factor improves the accuracy of AP estimates, we need a data set with independent estimates of the APs from spectroscopy. I use a set of 880 bright, nearby F,G,K dwarf stars observed with echelle spectroscopy for the Keck, Lick and AAT planet search programmes. Valenti \& Fischer~\citep{vf05} (hereafter VF05) estimated effective temperature, surface gravity, metallicity (\feh) plus various individual element abundances for these spectra. These estimates are based on Kurucz model atmospheres combined with various lines lists with empirical corrections. The stars all have near-solar metallicity (90\% with \feh\ between $-$0.4 and $+$0.45\,dex), are bright (90\% beween $V$\,=\,5.1--8.6\,mag) and have parallaxes from Hipparcos (90\% have distances ranging from 13--67\,pc). The full temperature range is 4707--6594\,K ($\log T$\,=\,3.673--3.819\,dex) with distribution shown in Fig.~\ref{v2dat_teff_histogram}.  I only use these effective temperature estimates (derived only from the spectrum) in what follows.\footnote{VF05 also estimate luminosity, radius and mass using in addition parallax and V-band photometry, but these are not used here.} Individual internal errors are about 50\,K (1$\sigma$).  I obtained the corresponding broad band $BVJHK$ photometry via the Hipparcos identifiers from Simbad ($JHK$ comes from 2MASS).\footnote{The full VF05 sample comprises 1040 stars, but only 885 had full five-band photometry in Simbad, and five had highly deviant photometry, leaving 880.}

VF05 do not provide estimates of the line-of-sight extinction, but given their proximity it is reasonable to assume it is negligible for most stars. Yet in order to train and validate the model, we need a sample with variance in $\a0$, which I introduce via simulations.  Ideally one would take a sufficiently high resolution SED for each star and calculate the extinction in each band for a range of $\a0$ (at fixed $\r0$) using equation~\ref{eqn:extint}. This, however, would be difficult to do reliably with the original VF05 data because of the problems of flux calibrating echelle spectra. I instead represent the SED with a synthetic spectrum at the known effective temperature of each star.  In fact, because these are broad band filters, the extinction in a band varies smoothly with $T$, so we do not need a synthetic SED for every unique $T$.  It is instead sufficient to calculate the integral on a discrete grid of six values of $T$ (I use SEDs from MARCS models) for a range of $\a0$ and to then make a series of one-dimensional quadratic fits, $A_b = y_b(T ; \a0)$. These functions gives the extinction in band $b$, $A_b$, as a smooth function of $T$ for each $\a0$.  Extinction is then applied to the original $BVJHK$ photometry by adding the corresponding $A_b$.
I do this for six values of $\a0$ (0.0 to 2.5\, mag in steps of 0.5\,mag) and apply it to the 880 stars ($T$ values) in the catalogue.  This produces an extended catalogue of 5280 stars showing variance in $\a0$ and $T$ with consistent APs, photometry and astrometry.

The parallaxes are not changed by this procedure.  This grid of stars is therefore not characteristic of the solar neighbourhood, because I have introduced numerous fainter, extincted stars with the same $T$ and distance as each unextincted star.  This is entirely appropriate, because the goal is only to achieve variance in $T$ and $\a0$ for fitting the forward model. 
 
Note that the MARCS models have only been used to simulate the {\em changes} in the colours due to extinction. (The temperature variation is still from the original data.)
This makes the data set relatively insensitive to any inaccuracies in the MARCS model's predictions of the colours at the six temperatures used in the fits. 
Compared to the calibrations of Worthey \& Lee~\citep{wl06},
Vallenari (private communication) 
finds a systematic offset in the predicted MARCS colours of $-$0.05\,mag in $B-V$ at 4770\,K, which corresponds to a systematic offset in $T$ of $+$50\,K or 1\%. (The systematic offsets in the redder colours are smaller.)
Although these systematics only affect the non-zero extinction data, they will slightly bias the whole forward model to give slightly higher $T$ estimates relative to the Worthey \& Lee calibration. Overall, the effective temperature scale predicted by the forward model is an amalgam of the VF05 and MARCS systems.

In the method I reduce the five-band photometry to the four colours $\vecp = (B-V, V-J, J-H, H-K)$.  While the five band measurements are independent, the four colours are not, so we must include the covariant (off-diagonal) terms into the covariance matrix in equation~\ref{eqn:likelihood} (see appendix).  As these are bright stars their uncertainties are not set by photon statistics but rather by object-independent sources. I therefore adopt an estimate of the photometric uncertainty of 0.021\,mag in each band, corresponding to $\sigma_p=0.03$\,mag in each colour.

\subsection{Forward model fitting and likelihood calculation}\label{sect:fmfit}

\begin{figure}
\begin{center}
\hspace*{-1.2em}
\includegraphics[width=0.50\textwidth]{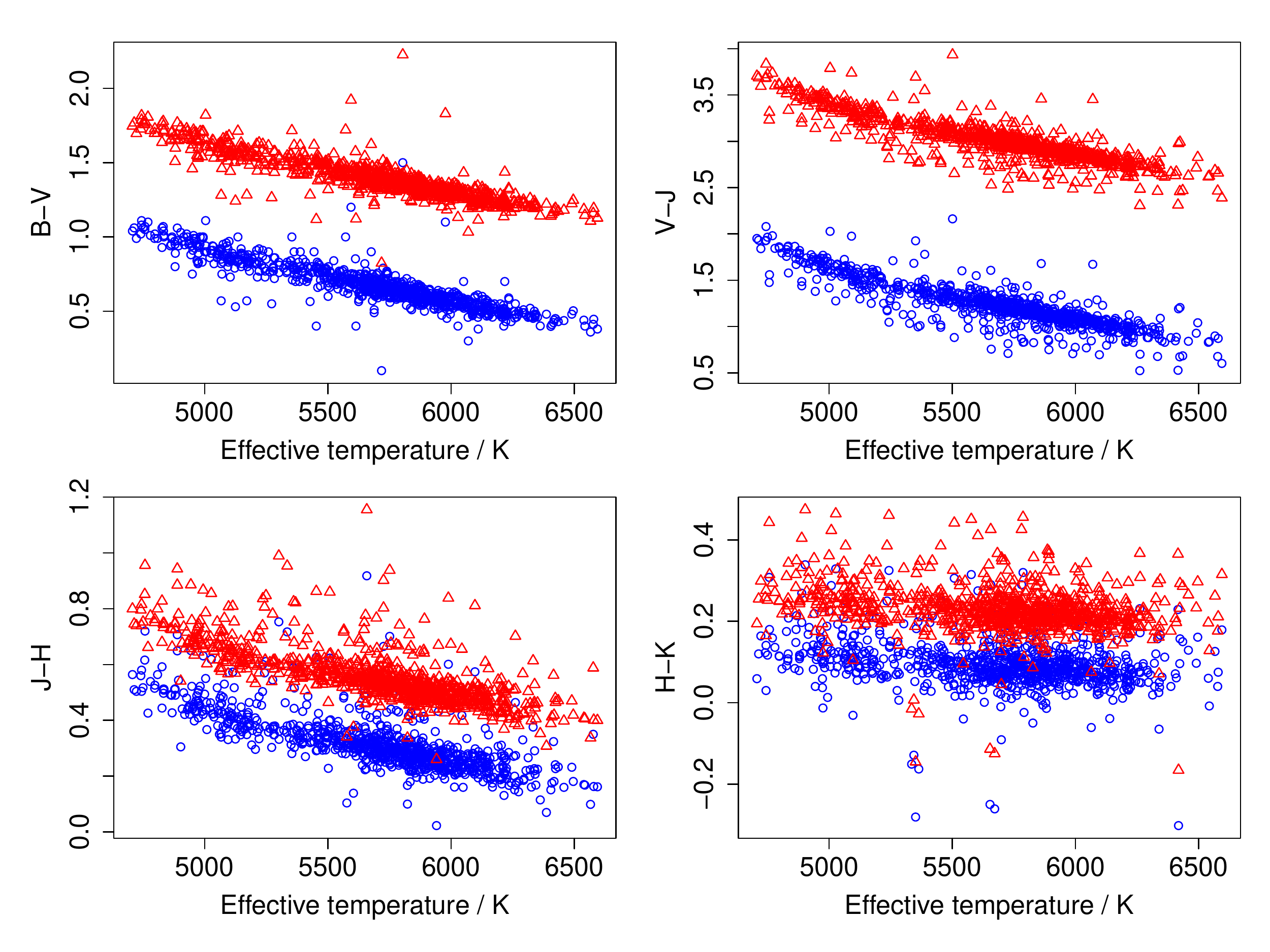}
\caption[]{Colour--temperature relation for the stars used to fit and test the model, for $\a0$\,=\,0\,mag (blue circles) and $\a0$\,=\,2.5\,mag (red triangles).
\label{v2ext_col_vs_teff}}
\end{center}
\end{figure}

The forward model defined in section~\ref{sect:fm} was fit to each of the four colours as a function of $\a0$ and $\log T$ using a quarter of the 5280 stars in the extended catalogue selected at random. A two-dimensional smoothing spline with 10 degrees of freedom gave a good fit in both directions.\footnote{The extended catalogue was initially calculated on a denser grid of $\a0$, but this did not significantly improve the forward modelling accuracy.}
With either AP fixed, the four colours vary almost linearly with the other AP (see Fig.~\ref{v2ext_col_vs_teff}).  The sensitivities of the four colours with respect to temperature at zero extinction (the gradients $|\partial p_i/\partial \log T|_{\a0=0}$) are approximately $(-4.4, -7.3, -2.4, -0.4)$ mag/dex for $(B-V, V-J, J-H, H-K)$ respectively. (Multiply these numbers by $\log1.1=0.041$ to get the approximate colour change due to increasing $T$ by 10\%, for example.)
The sensitivities of these colours with respect to $\a0$ at $T$\,=\,5500\,K are $(0.29, 0.71, 0.10, 0.05)$ mag/mag.  The scatter in the colours at fixed $\a0$ and $T$ in 100\,K temperature wide bins is typically 0.05--0.1\,mag for $B-V$, $J-H$ and $H-K$, but larger in $V-J$ (0.1--0.15\,mag) and at around $T$\,=\,5800\,K. This scatter is partly due to photometric errors but is mostly a result of cosmic variance. This obviously limits the accuracy of AP estimation with these data.  For example, if we knew the extinction, then
estimating $T$ using just the $V-J$ colour with an error of 0.1\,mag would translate into an error in $\log T$ of 0.014\,dex or 3\%.
Although the use of four colours permits better estimates, neither AP is known a priori and, as we shall see, there is an intrinsic degeneracy between these APs with these colours.


Once fit, the forward model is used to to calculate the likelihood $P(\vecp | \a0,T)$ (section~\ref{sect:likelihood}) for the remaining 3/4 of the stars in the extended catalogue.
A common way of sampling probability distributions is with Monte Carlo methods, but this is inefficient for a two-dimensional parameter space. I instead sample it on a dense, regular grid with $\a0$ varying from 0.0 to 3.5\,mag in steps of 0.05\,mag and $\log T$ varying from $\log 4300$ to $\log 7000$ in steps of 0.005\,dex, yielding a grid of $71 \times 43 = 3053$ points. I will refer to this as the {\em d-grid}.
The limits on $T$ at which we calculate this grid are set by the temperature range in the catalogue and avoiding wanting to extrapolate the forward model more than about 400\,K in $T$.

\subsection{p-model results}

\begin{figure*} 
\begin{center}
\includegraphics[width=0.90\textwidth]{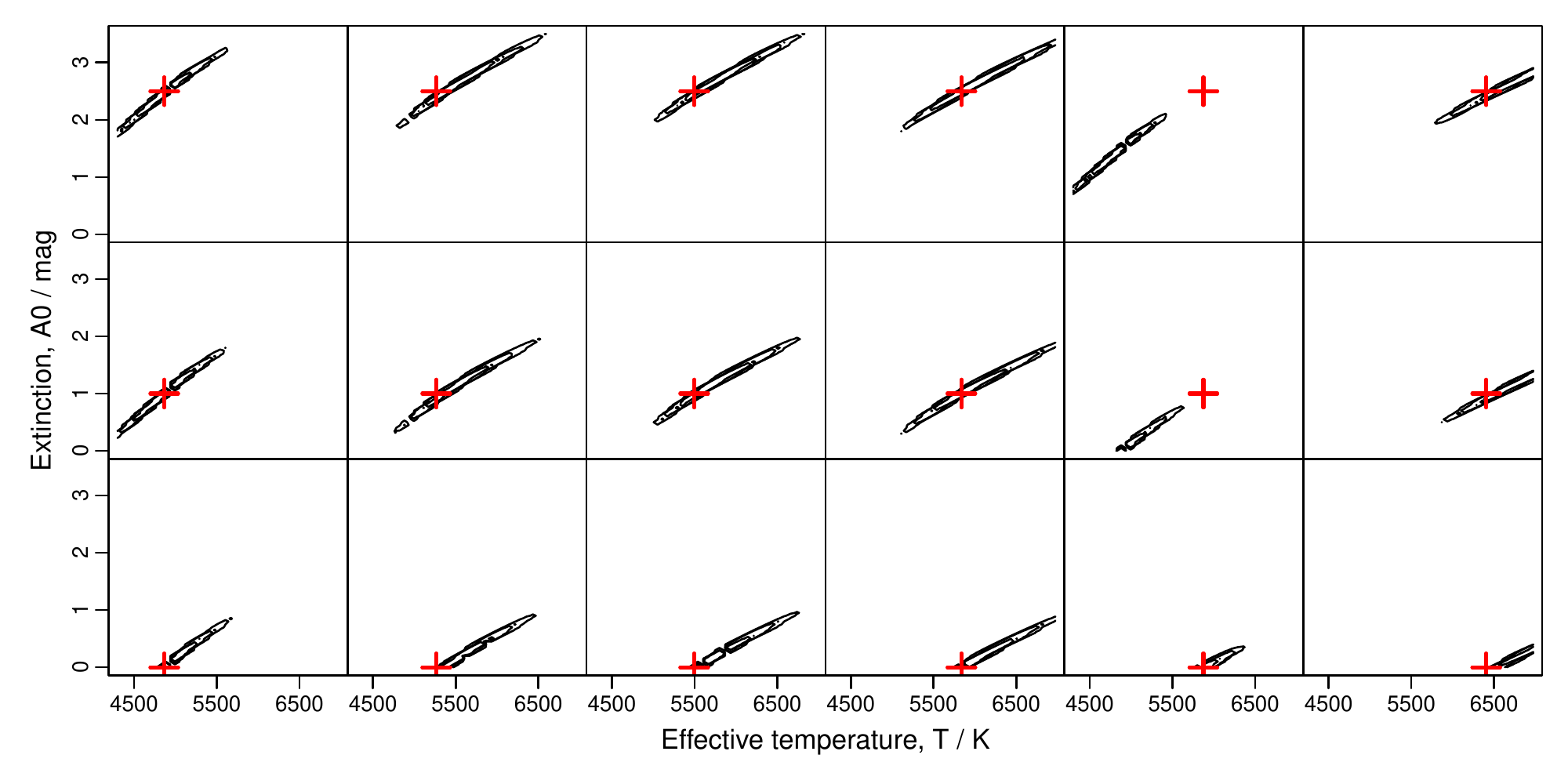}
\caption[]{Posterior probability density function (PDF) from the p-model over 18 stars with 6 different true temperatures (columns) and three different true extinctions (rows) applied to the extended catalogue. Three contours are shown for each star, each enclosing 90\%, 99\% and 99.9\% of the total posterior probability. For comparison, the true AP values are shown with the red cross. 
\label{test15_pdfat_contour_pmodel}}
\end{center}
\end{figure*}

The p-model is the posterior PDF (equation~\ref{eqn:pdfat}) over the two APs assuming uniform priors $\a0$ and $T$, i.e.\ not using the HRD/q factor. This posterior, $P(\a0,T | \vecp)$, can be summarized in a two-dimensional contour plot and is shown in Fig.~\ref{test15_pdfat_contour_pmodel} for some example stars, whereby the contours contain 90\%, 99\% and 99.9\% of the total posterior probability (found by integrating the PDF).

The most obvious feature common to all these plots (and for the overwhelming majority of stars in the extended catalogue) is the strong $\a0$--$T$ degeneracy revealed by the long, narrow contours. {\em This degeneracy is intrinsic to the data and not a feature of the method.} This is because for given colours corresponding to some nominal $\a0$ and $T$, we will get the same colours by increasing both APs or by decreasing both APs by a certain amount.  We would even have the degeneracy with negligible photometric errors, although in that case the degenerate region would be smaller (narrower region). The contours show a slight curvature in these plots against $T$. In the original $\log T$ space they are straighter with a slope $d\a0/d\log T \simeq$\,10\,mag / dex.

Not surprisingly, the APs cannot be estimated very accurately. In most of these plots the 90\% contour extends over approximately 1000\,K in effective temperature and over 1\,mag in extinction. Using just these data we can neither estimate the parameters more precisely nor remove the degeneracy. That can only be done using additional (prior) information, e.g.\ if we had other data which indicated that the extinction was small.

\begin{figure*} 
\begin{center}
\includegraphics[width=0.90\textwidth]{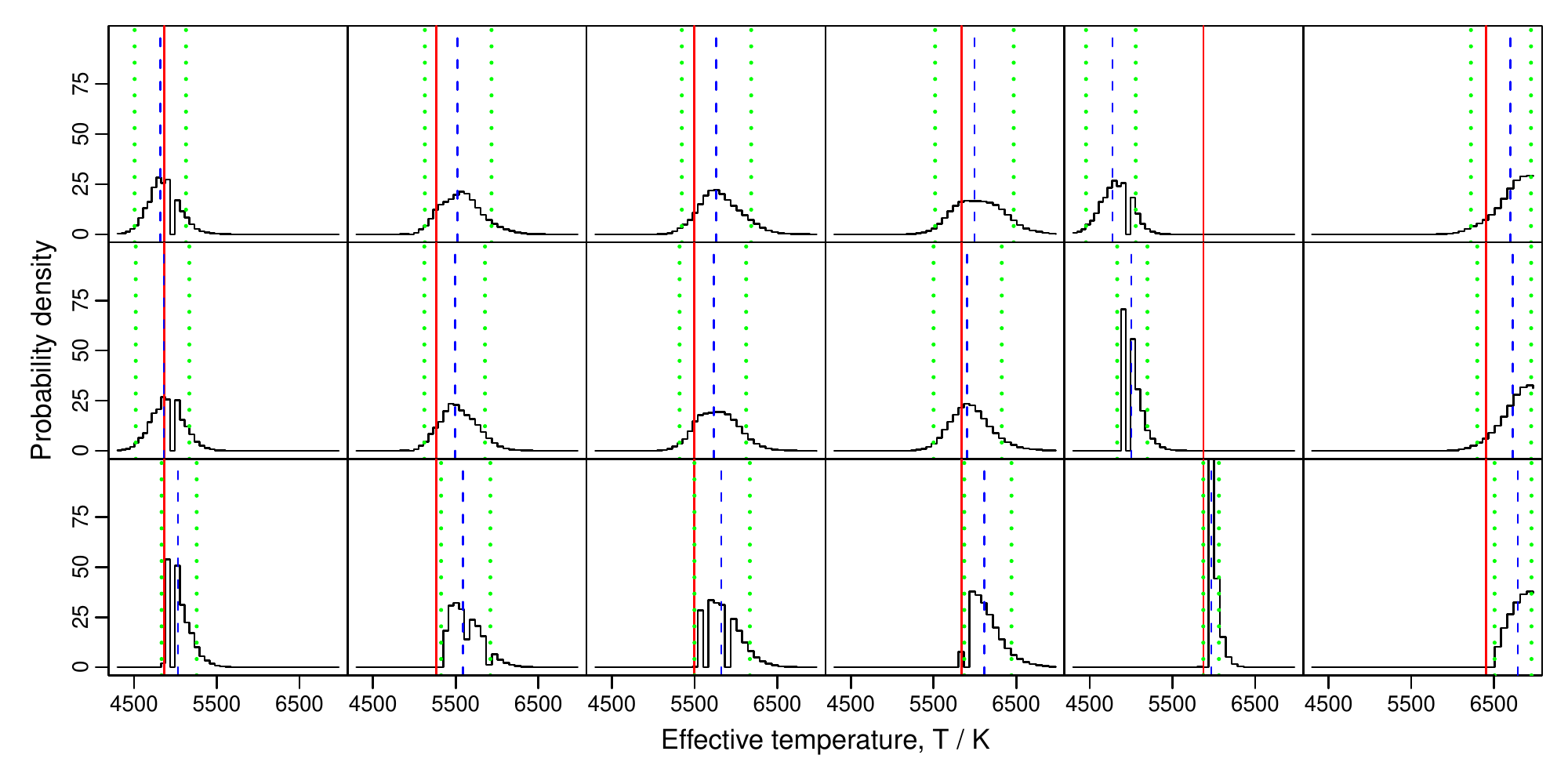}
\caption[]{One-dimensional posterior PDFs (p-model) over temperature for the stars shown in Fig.~\ref{test15_pdfat_contour_pmodel} obtained by marginalizing those two-dimensional PDFs over $\a0$. The probability density is properly normalized (integral is unity).
The dashed blue line is the mean of the 1D PDF, the green dotted lines denote the 90\% confidence interval. The ``true'' temperature (from VF05) is shown with a solid red line.
\label{test15_pdfat_teff_contour_pmodel}}
\end{center}
\end{figure*}

From a Bayesian perspective the full PDF is the final answer to the inference of the APs. But for surveys of many objects we want to summarize this with a best estimate plus a confidence interval. As the PDFs are predominantly unimodal and symmetric, I summarize using the mean and 90\% confidence interval for each AP (plus the correlation coefficient, the slope of the major axis of the contours).  These may be obtained by marginalizing the PDF over each AP (equation~\ref{eqn:pdfatmarg}). The result of the marginalization over $\a0$ is shown in Fig.~\ref{test15_pdfat_teff_contour_pmodel}. The 90\% confidence interval is the range which contains the central 90\% of the probability. It is not necessarily symmetric about the mean. For a Gaussian distribution this interval is the $\pm1.64\sigma$ interval, so the equivalent ``$1\sigma$ error'' is 3.3 times smaller.

The posterior PDF is calculated on and normalized over a discrete, albeit fine, grid (the d-grid), with the 90\% confidence interval calculated via discrete integration with linear interpolation.  This finite-sized grid truncates the PDF when it extends to the edge of the grid.  Some adjustments are therefore made in order to report sensible confidence intervals when the PDF peaks very strongly at the edge of the grid. The constraint at $\a0$\,=\,0 is a natural one because extinction cannot be negative, but the other three constraints are artificial. We see in the right column of Fig.~\ref{test15_pdfat_teff_contour_pmodel} an example of how this edge peaking translates into contraints on the inferred mean and confidence interval. This actually only occurs for a small fraction of the stars in the extended catalogue. It could be avoided entirely if we chose to extend the $\a0$ range or to extrapolate the forward model further over $T$.

In the fifth column of Fig.~\ref{test15_pdfat_contour_pmodel} I show cases where the true APs lie outside of the 90\% confidence interval of the p-model estimation. This could be a consequence of (a) cosmic scatter (other parameters) affecting the data, (b) large errors in the data which are not reflected by the estimated uncertainties, $\{\sigma_{p_i}\}$, in the likelihood model, (c) an error in the VF05 $T$ assignment, or a combination of these.
There also exist cases where the contours lie at higher $\av$ and $T$ than the ``true'' APs. This could also be an indication that the original star had non-zero extinction.

\begin{figure}
\begin{center}
\includegraphics[width=0.50\textwidth]{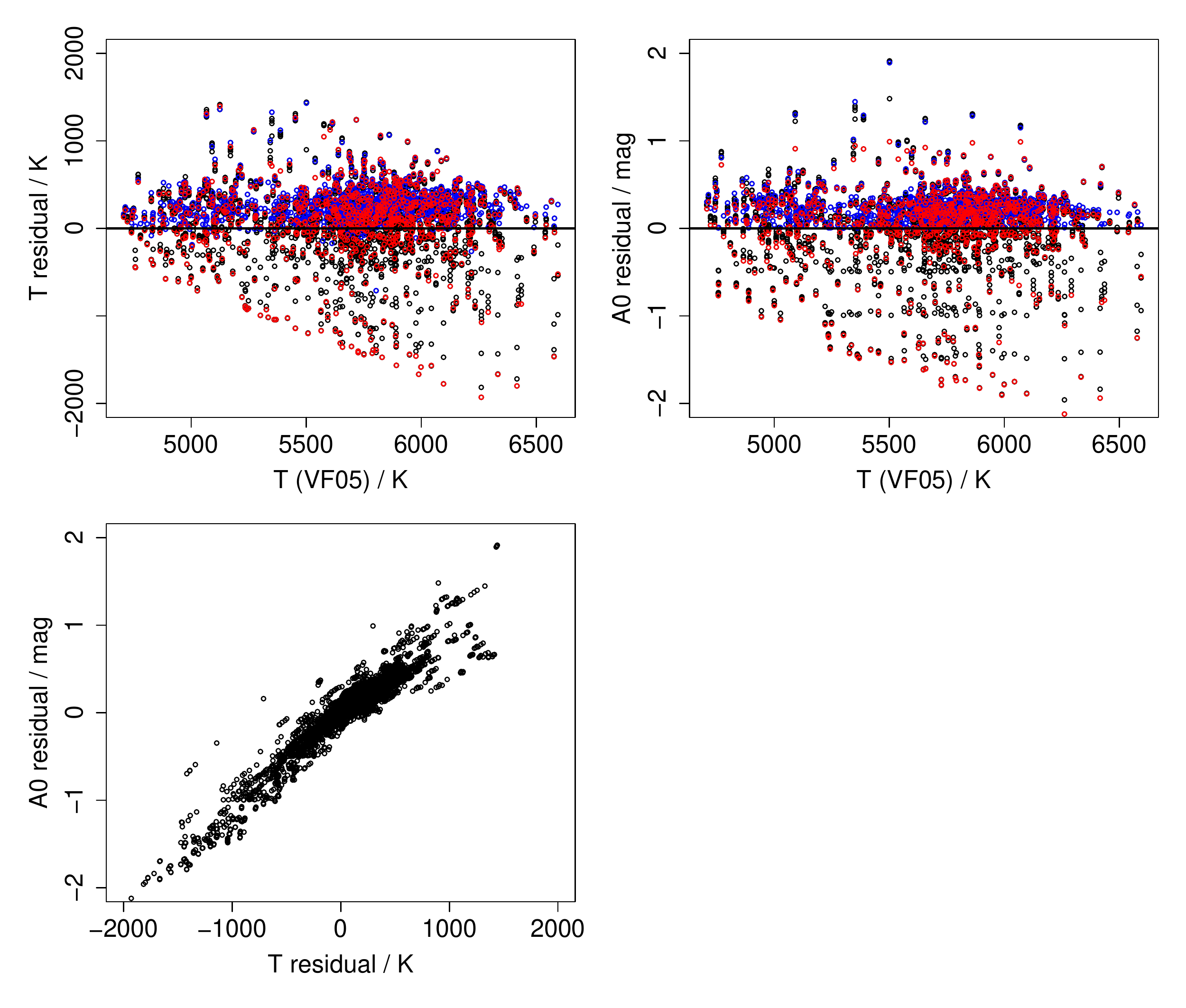}
\caption[]{Parameter residuals for the p-model applied to the extended catalogue. The residual is defined as p-model prediction minus true value.
Points with true $\a0$ equal to 0.0 and 2.5\,mag are coloured blue and red respectively.
\label{test15_apresid_pmodel}}
\end{center}
\end{figure}

The overall {\em accuracy} of the p-model results is shown in Fig.~\ref{test15_apresid_pmodel}, which plots the AP residuals,
$\delta \phi \!=\! \phi_{\rm inferred} \!-\! \phi_{\rm true}$,
against $\log T$ and each other. 
There is no particular trend in the residuals, other than evidence of the d-grid limits in the
form of diagonal trends at the edges of the upper two plots. The third panel shows the not surprising result that the residuals are correlated. I summarize the accuracy using the 
mean absolute residual, $\overline{|\delta \phi|}$ (abbreviated as \mar), and the mean residual, $\overline{\delta \phi}$ (abbreviated as \mr), which are summaries of the random errors and systematic errors respectively. \mar\ is 0.024\,dex (5\%) for $\log T$ (300\,K for a solar-type star of 5800\,K) and 0.29\,mag for $\a0$. (Multiply these by 1.25 to get the $1\sigma$ for a Gaussian distribution.) The systematic errors, apparent from the plots, are 0.008\,dex (2\%) and 0.08\,mag respectively.

The two main sources of the random errors are (1) the photometric errors, plus the inevitable non-Gaussianity of the error distribution, and (2) the intrinsic degeneracy, or the fact that these four colours alone are insufficient even in principle to estimate both APs exactly. The source of the small systematic errors is less clear. Both train and test data are drawn from the same sample, so it is not an issue of a simple data mismatch.
The systematic may be a consequence of the slight biasing introduced by the use of the MARCS models (section~\ref{sect:fitvald}), which could have introduced an inconsistency in the modelled variation of the colours over the APs.

\begin{figure}
\begin{center}
\hspace*{-1em}
\includegraphics[width=0.50\textwidth]{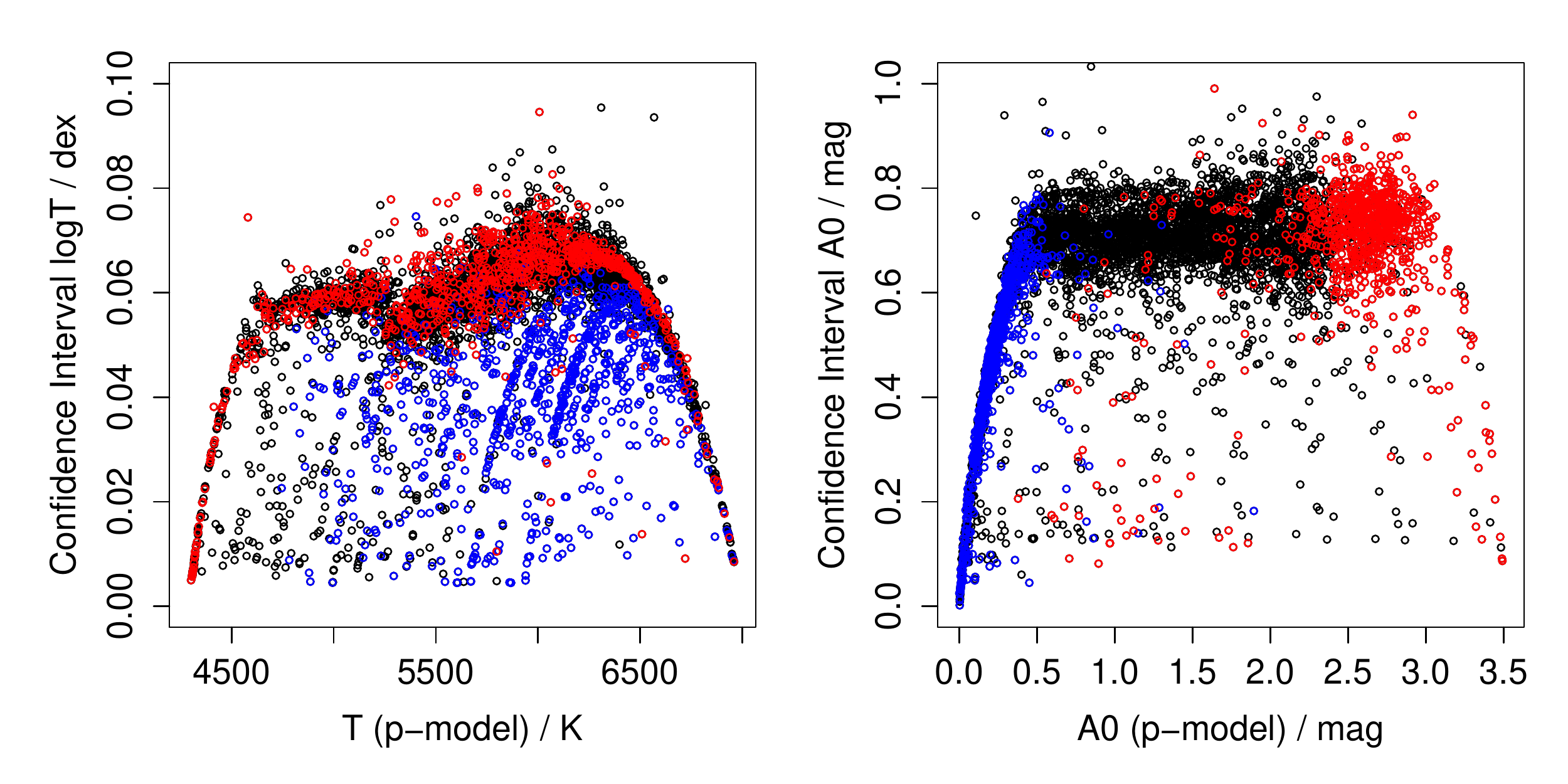}
\caption[]{The 90\% confidence intervals of the APs inferred from the extended catalogue with the p-model, plotted against the estimated mean AP. To get the equivalent Gaussian $1\sigma$ error mutiply by 0.3. Points with true $\a0$ equal to 0.0 and 2.5\,mag are coloured blue and red respectively. 
\label{test15_CIs_pmodel}}
\end{center}
\end{figure}

The {\em precision} of the AP estimates (how good we think they are in the absence of the truth) is measured by the 90\% confidence intervals.  These are typically 0.07\,dex (16\%) in $\log T$ and 0.7\,dex in $\a0$, with relatively little dependence on the APs except at the edge of the d-grid (see Fig.~\ref{test15_CIs_pmodel}).

Let us now introduce an HRD prior and a model for the $q$ constraint and examine how these alter the inference.

\subsection{Model for the HRD prior}\label{sect:hrdpriormodel}

\begin{figure}
\begin{center}
\includegraphics[width=0.42\textwidth]{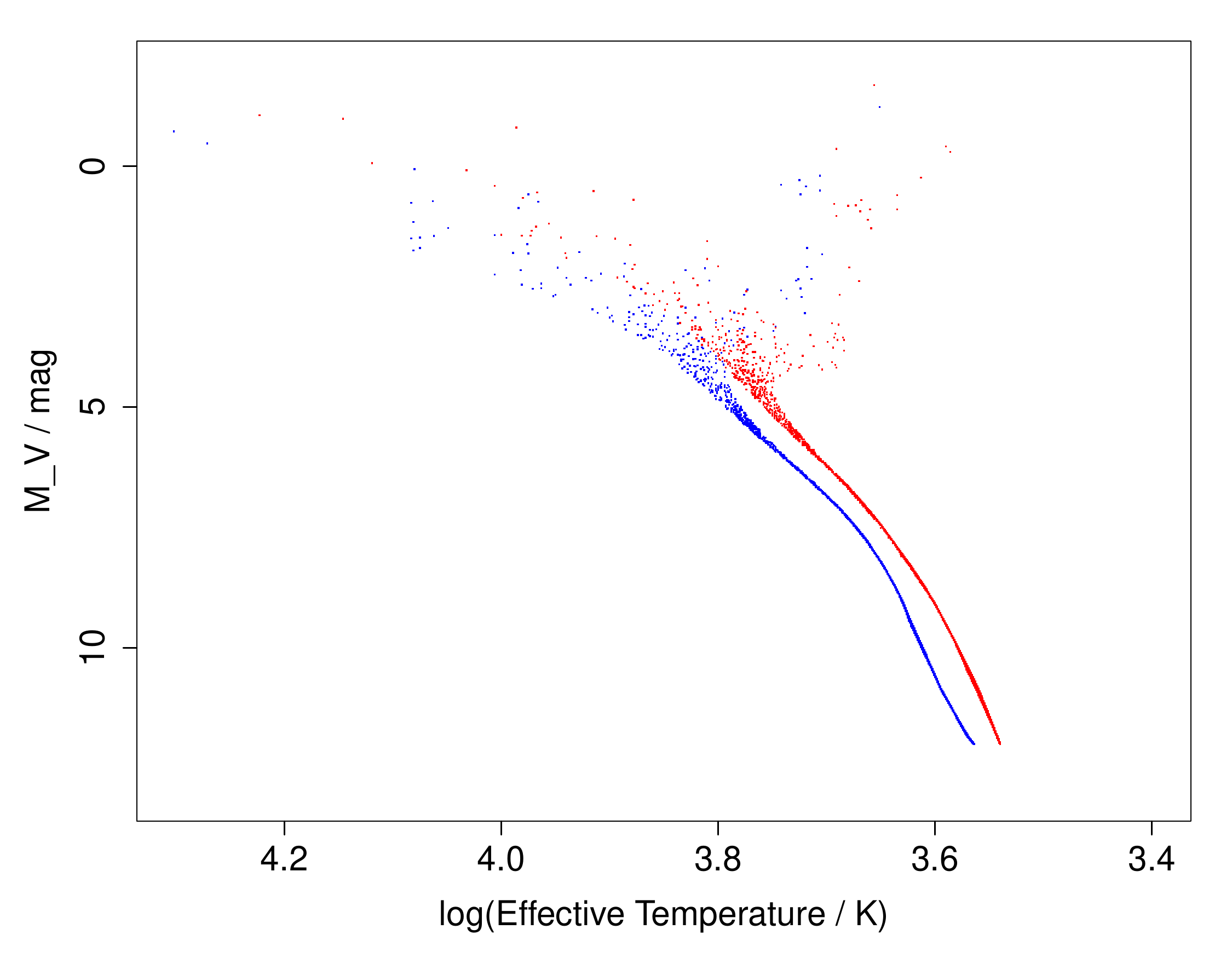}
\caption[]{Stellar evolution simulations from Vallenari et al.~\citep{v10} used to build the HRD prior. The stellar locus shown in red points (main sequence to the right) is for $Z=0.019$ and the one in blue points (to the left) is for $Z=0.0019$. Only every tenth star 
and only the lower mass stars in the simulations are shown.
\label{hrd_vband_lowmass_z019_z_0019_forpaper}}
\end{center}
\end{figure}

\begin{figure}
\begin{center}
\includegraphics[width=0.50\textwidth]{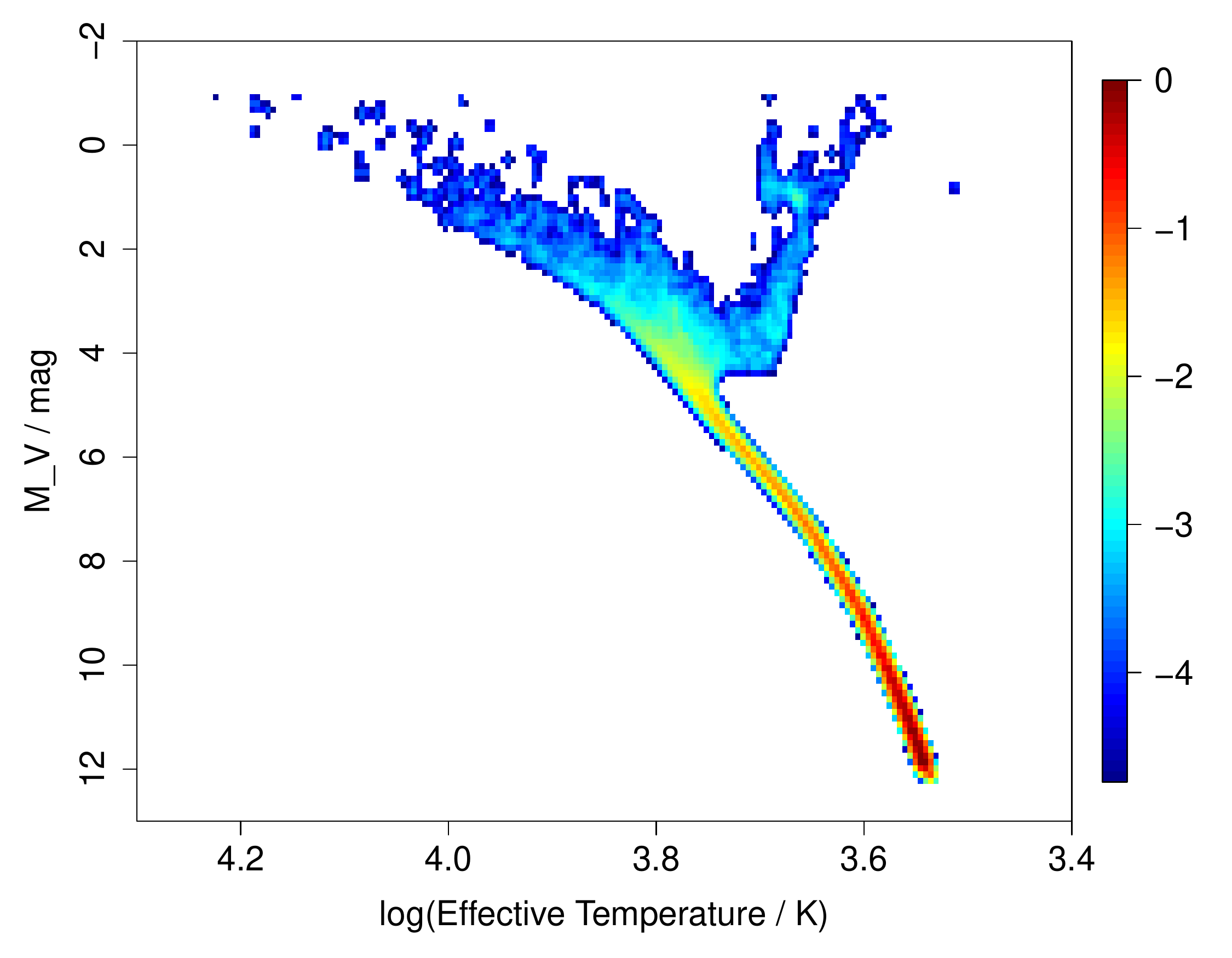}
\caption[]{HRD prior. The colour scale shows $\log P(\mv, T)$ normalized for the purpose of illustration to have zero at its maximum. Unoccupied areas are shown in white.
\label{hrdmap_vband_comb_z019_8_joinB_forpaper}}
\end{center}
\end{figure}

The HRD prior, $P(\mv, T)$, is the prior probability over $\mv$ and $T$ based on our knowledge of stellar evolution and stellar populations. We have some choice to exercise in what evolution/population we represent. For the sake of this article I use the results of a complete evolution of a simulated stellar population generated by the code of Vallenari et al.~\citep{v10} based on the evolutionary models of Bertelli et al.~\citep{b08}.
The population comprises 200\,000 stars drawn from a Salpeter IMF with initial masses ranging from 0.2--107\,\Msol\ (although 99\% of stars have masses below 1.3\,\Msol).  A constant star formation rate over a period of 13.7\,Gyr is applied and the stars are evolved independently. Solar metallicity, $Z$\,=\,0.0190--0.0193, is used for all stars.
The final $\mv$ and $T$ of these stars are calculated and used to place them in the HRD.
A subset is shown as the red sequence in Fig.~\ref{hrd_vband_lowmass_z019_z_0019_forpaper}. This represents the HRD prior as a set of delta functions. This could be used as is, but in order to accommodate some cosmic variation and overcome the sparseness of the simulation in some parts of the HRD, I smooth it using two-dimensional kernel density estimation.  I use a Gaussian kernel with kernel widths of 0.0025\,dex in $\log T$ and 0.0625\,mag in $\mv$ (and zero covariance) and calculate the density on a 200$\times$200 grid. Normalizing so that the total integral is unity gives the prior $P(\mv, T)$. This is shown using a log density scale in Fig.~\ref{hrdmap_vband_comb_z019_8_joinB_forpaper}. Note the very large range in densities (almost five orders of magnitude). 

This HRD prior is certainly not perfect. Its main defect is that it considers only a single metallicity. While that is not an issue for the solar metallicity VF05 sample, it is a limitation for the later application to the Hipparcos/2MASS sample in section~\ref{sect:hip2mass}. Metallicity significantly affects the position of the stellar loci in the HRD, as we can see in Fig.~\ref{hrd_vband_lowmass_z019_z_0019_forpaper}.  Comparison of the main sequences in this figure with that in the smoothed HRD prior shows that quite a liberal smoothing has been applied, making this a rather weak prior which could be interpreted as a small prior range on metallicity of order $\pm 0.5$\,dex. I shall later investigate the impact of the HRD metallicity empirically.

\subsection{Model for the $q$ constraint}\label{sect:qconstmodel}

The $q$ constraint, $P(q | \mv, \a0, T)$, introduced in section~\ref{sect:qconstraint}, is a one-dimensional PDF over
$q - (\mv + \av - 5)$ with zero mean, where $\av$ is a function of $\a0$ and $T$.
Although, from equation~\ref{eqn:q}, $q$ will not strictly have a Gaussian distribution when $V$ and $\varpi$ have Gaussian distributions, we can still approximate it as such provided we calculate its variance, $\sigma^2_q$, correctly.  This must take into account the observational correlation between magnitude and parallax. 
Writing the variances of the photometric noise and the parallax noise as
$\sigma^2_V$ and $\sigma^2_{\varpi}$ respectively,
it follows from the definition of $q$ and a general result of covariances
(equation~\ref{eqn:covsum}) that
\begin{equation}
\sigma_q(V,\varpi)^2 \,=\, \left(\frac{5}{\ln 10}\frac{\sigma_{\varpi}}{\varpi}\right)^2 \ + \ \sigma_V^2 \ + \ 2\left(\frac{5}{\ln 10}\frac{\sigma_{\varpi}}{\varpi}\right)\sigma_V\rho(V,\log\varpi) 
\label{noisemod} 
\end{equation}
where $\rho$ is the correlation coefficient.

Calculated over the original catalogue of 880 objects $\rho(V,\log\varpi)$\,=\,$-$0.62, the value I adopt for all calculations.
A typical star in the original catalogue has $V$\,=\,7\,mag, $\sigma_V$\,=\,0.02\,mag, $\varpi$\,=\,30\,mas and $\sigma_{\varpi}$\,=\,1\,mas which gives $\sigma_q$\,=\,$\sqrt{0.0052+0.0004-0.0018}=0.06$\,mag (note that it is dominated by the parallax error). Even the largest value of $\sigma_q$ (for a star with a small parallax and large parallax error) is just 0.62\,mag (90\% have $\sigma_q<0.13$).

How does the $q$ constraint work?  Having measured $q$, it constrains $\mv + \a0$ to within a range of order $\pm \sigma_q$ about this measurement. For a given value of $\a0$ this limits the range of $\mv$ and hence, via the HRD prior, the range of $T$. In other words, there is a finite-width probability distribution over $T$ for a given $\a0$. The ensemble of all such distributions is the two-dimensional PDF over $(\a0,T)$ for given $q$, which is just $P(\a0,T | q)$ in equation~\ref{eqn:a07}.

One more component is required to use this in practice.
The $q$ constraint is defined in terms of the extinction in a band, $\av$, whereas our goal is to get a posterior PDF over the extinction parameter $\a0$  (see section~\ref{sect:extinction}). The two are closely related however, and it is reasonable to assume that to a good approximation
\begin{equation}
\av \,=\, \a0 + y(\a0, T)
\label{eqn:ava0conv}
\end{equation}
where $y$ is a smooth, two-dimensional function.  I fit this using simulated $V$-band photometry calculated from the same MARCS synthetic spectra used to create the extended catalogue (section~\ref{sect:datacon}). I use a regular grid in $T$ (6 values) and $\a0$ (21 values).
The variation is quite smooth and is fitted accurately with a thin-plate spline.
Because $\a0$ is defined at a monochromatic wavelength close to the centre of the $V$ band, 
$y(\a0, T)$ is a small (negative) correction to $\a0$. Its largest value is $-0.18$\,mag, occuring for the highest extinction ($\a0$\,=\,3.5\,mag) and lowest temperature (4000\,K) considered.

\subsection{pq-model results}

\begin{figure*} 
\begin{center}
\includegraphics[width=0.90\textwidth]{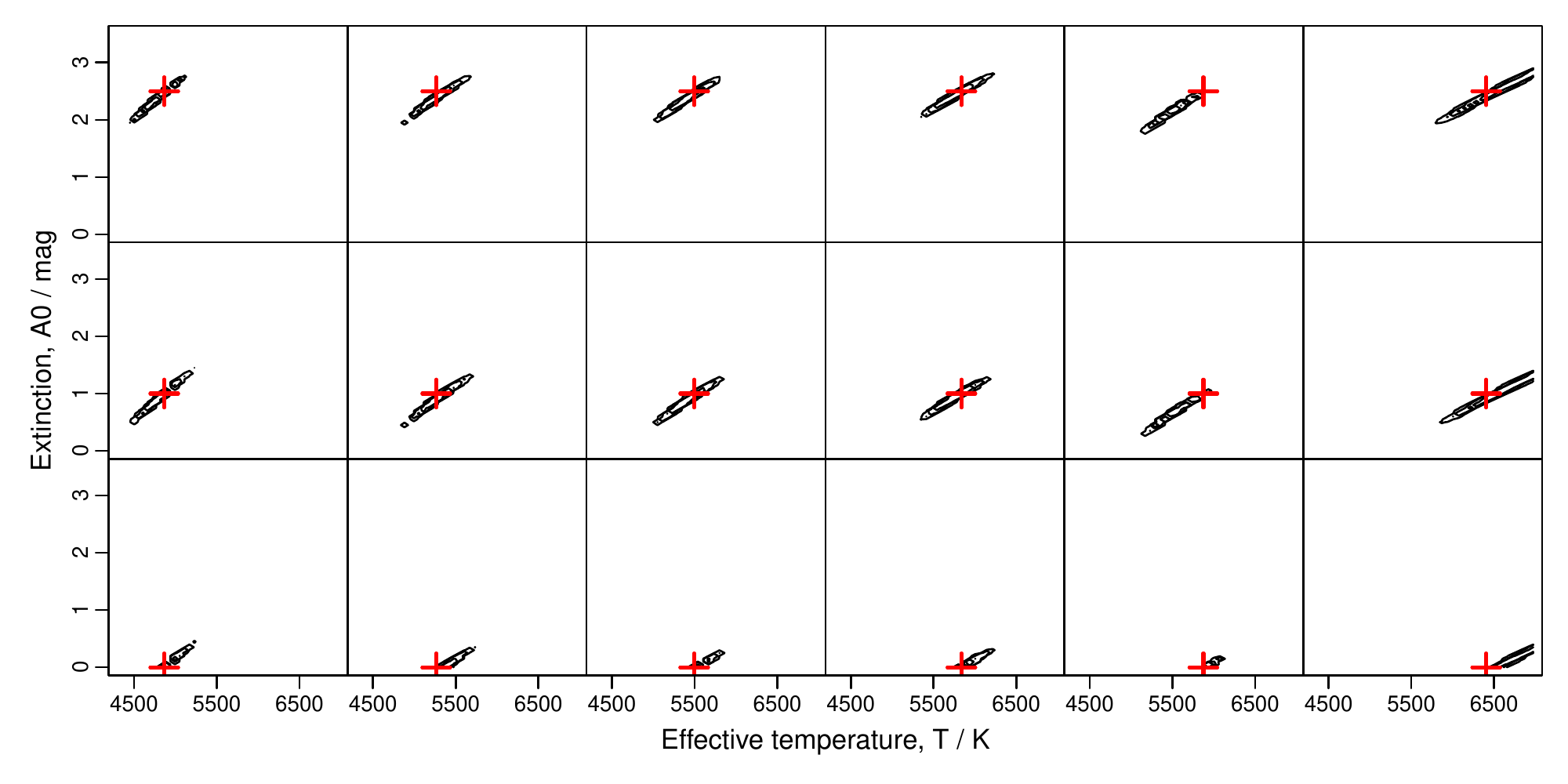}
\caption[]{Posterior probability distribution for the pq-model (i.e.\ including the HRD/q factor) for the same
stars as shown in Fig.~\ref{test15_pdfat_contour_pmodel}.
\label{test15_pdfat_contour_pqmodel}}
\end{center}
\end{figure*}

\begin{figure*} 
\begin{center}
\includegraphics[width=0.90\textwidth]{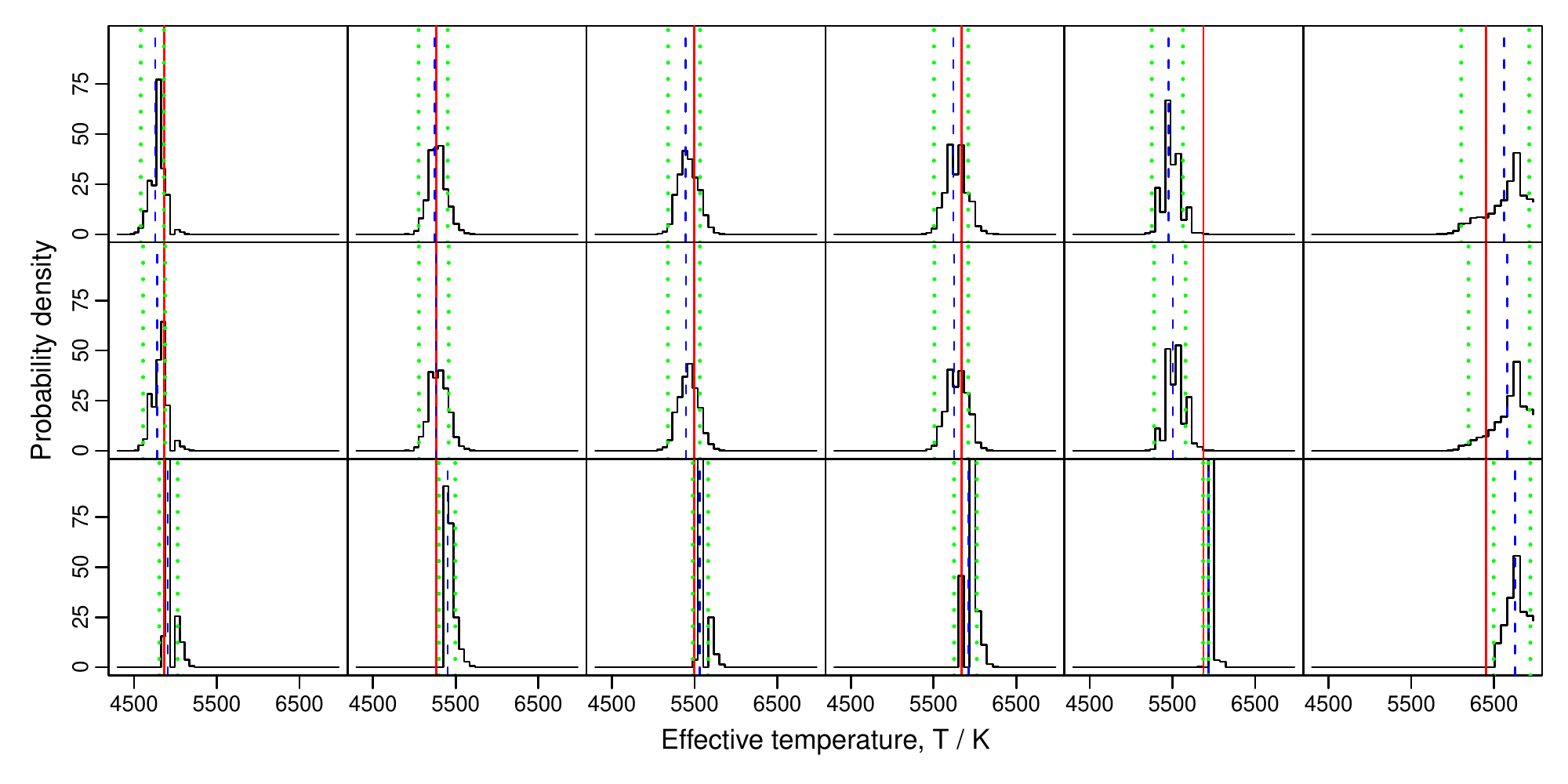}
\caption[]{One-dimensional posterior PDF (pq-model) over temperature for the stars shown in Fig.~\ref{test15_pdfat_contour_pqmodel} obtained by marginalizing those two-dimensional PDFs over $\a0$. 
The dashed blue line is the mean of the 1D PDF, the green dotted lines denote the 90\% confidence interval. The ``true'' temperature (from VF05) is shown with a solid red line.
\label{test15_pdfat_teff_contour_pqmodel}}
\end{center}
\end{figure*}

The pq-model was applied to the full extended catalogue using this HRD prior and $q$ model.  The resulting PDFs over the same stars as shown for the p-model in Fig.~\ref{test15_pdfat_contour_pmodel} are shown in Fig.~\ref{test15_pdfat_contour_pqmodel}. A  comparison shows that the introduction of the HRD/q factor has improved the precision (reduced the size of the contours). This can also be seen in the one-dimensional marginalized PDFs (Fig.~\ref{test15_pdfat_teff_contour_pqmodel}) as a reduction in width and increase in height of the distributions. The HRD/q factor has not removed the degeneracy, but it has reduced it.

\begin{figure}
\begin{center}
\includegraphics[width=0.50\textwidth]{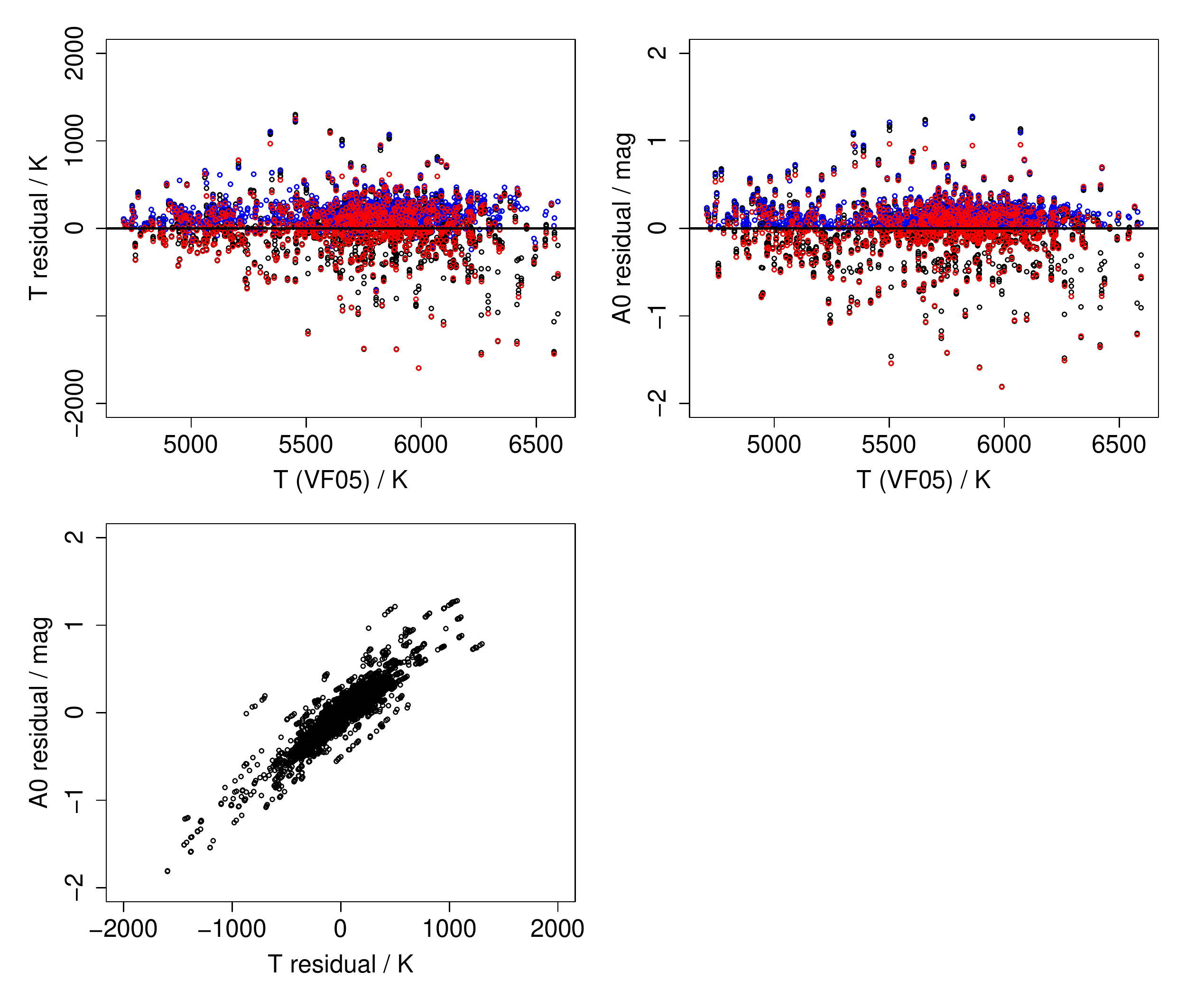}
\caption[]{AP residuals for the pq-model applied to the VF05 data. The residual is defined as pq-model prediction minus VF05 (``true'') value. Points with true $\a0$ equal to 0.0 and 2.5\,mag are coloured blue and red respectively.
\label{test15_apresid_pqmodel}}
\end{center}
\end{figure}

\begin{table}
\begin{center}
\caption{Performance of four parameter estimation methods on the extended catalogue. \mar\ (accuracy) is the mean absolute residual; \CI\ (precision) is the average 90\% confidence interval (multiply by 0.3 to get the equivalent Gaussian $1\sigma$ error); \Con\ (confidence) is the mean of the absolute ratio of residual to precision; \mr\ (systematic) is the mean residual; \semr\ (standard error in systematic) is \mar\,$\times 1.25/\sqrt{N}$, where $N$ is the number of objects in the test set.
\label{tab:perf_vf05}}
\begin{tabular}{llrrrrr}
\hline
Parameter  &  Method     & \mar\       & \CI\         & \Con\        & \mr\          &  \semr\ \\
\hline
$\log T$   & p-model    &  0.024     &  0.055     &  0.71          &  0.0075      &  0.0004 \\  
$\log T$   & pq-model  & 0.015      &  0.038     &  0.54          &  0.0037      &  0.0003 \\  
$\log T$   & ph-model  & 0.020      &  0.051     &  0.67          &  0.0026      &  0.0004 \\  
$\log T$   & \ilium\      &  0.026     &  0.071     &  0.36          &  0.0032      &  0.0005\\   
$\a0$       & p-model   & 0.29        &  0.63       &  0.88          &  0.075        &  0.005 \\
$\a0$       & pq-model  & 0.19       &  0.43       &  0.63          &  0.037       &   0.003 \\ 
$\a0$       & ph-model  & 0.25       &  0.59       &  0.85          &  0.020       &   0.004 \\ 
$\a0$       & \ilium\      & 0.38        &  0.92      &  0.41          &  0.079        &  0.008 \\
\hline
\end{tabular}
\end{center}
\end{table}

The critical question is whether the HRD/q factor has improved the {\em accuracy}. I again take the mean of the marginalized PDF distribution as the estimate of the AP.  The residuals for the pq-model are shown in Fig.~\ref{test15_apresid_pqmodel} and may be compared with the p-model residuals from before. We immediately see that the pq-model has the better performance in terms of both random and systematic errors.  The third column of Table~\ref{tab:perf_vf05} summarizes the random errors with the mean absolute residuals.  There is a marked improvement with the pq-model compared to the p-model: by 39\% in $T$ and 33\% in $\a0$ (measured as $[\epsilon_{\rm mar}^{p} - \epsilon_{\rm mar}^{pq}]/\epsilon_{\rm mar}^{p}$).  
The overall performance of 3.5\% (\mar) in $T$, which is 200\,K at 5800\,K, is quite good considering the limited data on which this is based, in particular when we recall that the intrinsic scatter in the photometry limits the photometric-only estimates to a similar value (as discussed in section~\ref{sect:fmfit}).
At 5800\,K this error corresponds to a Gaussian 1$\sigma$ of 250\,K which may compared to the internal errors on the training data of 50\,K.

Table~\ref{tab:perf_vf05} also quantifies the improvement in the precision (smaller \CI). Interestingly, the pq-model is more conservative than the p-model. This is measured by the {\em confidence}, the ratio of accuracy to precision, defined here as \Con\,=\,$\langle |\delta \phi_i|/ {\rm CI}_i \rangle$. Precision has not improved as much as the accuracy, so the pq-model has slightly lower confidence.

\begin{figure}
\begin{center}
\hspace*{-0.75em}
\includegraphics[width=0.50\textwidth]{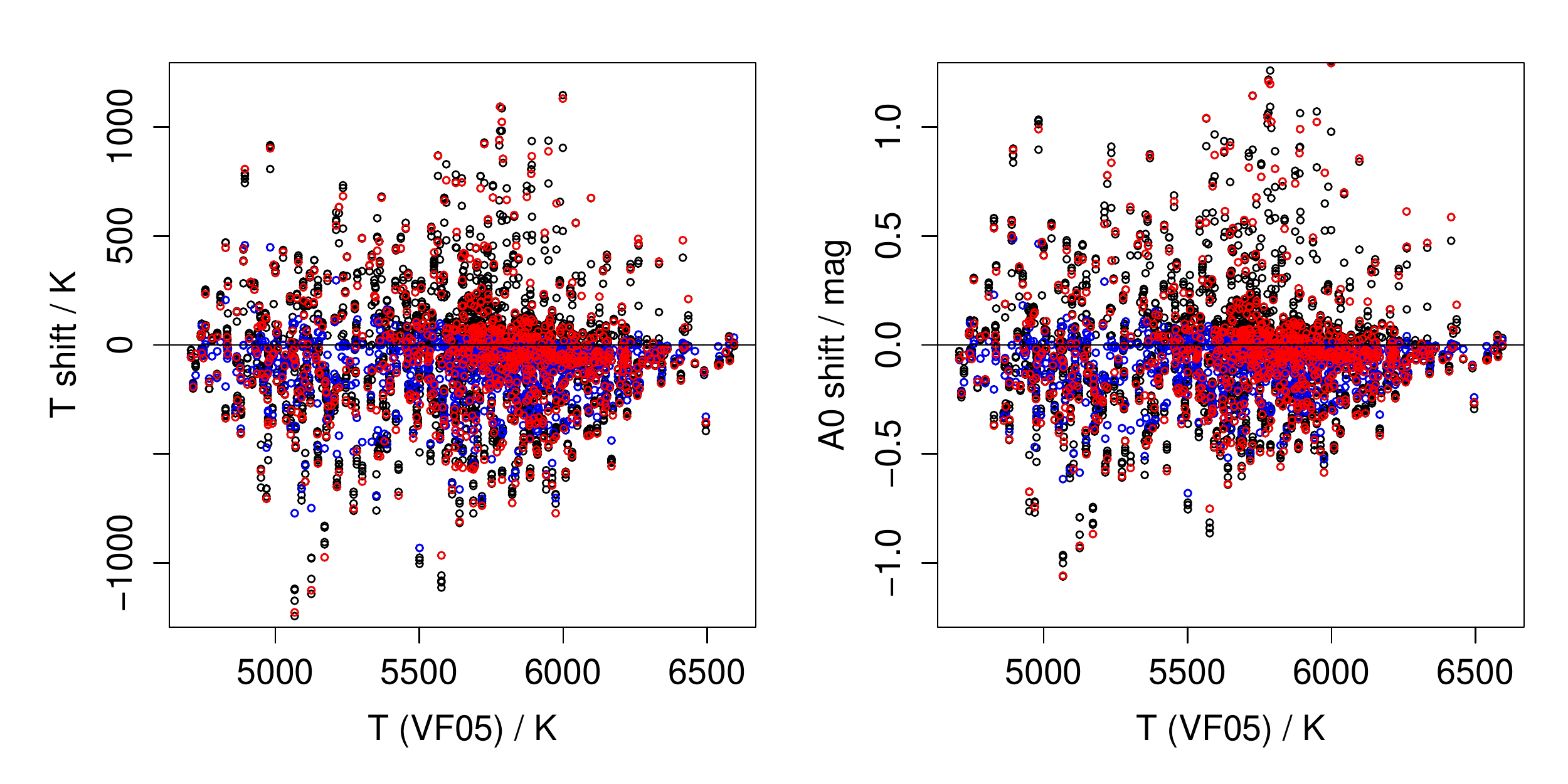}
\caption[]{Shift in mean estimated AP (pq-model minus p-model), plotted against the true (VF05) temperature.
Points with true $\a0$ equal to 0.0 and 2.5\,mag are coloured blue and red respectively.
\label{test15_apeshifts}}
\end{center}
\end{figure}

The HRD/q factor improves AP estimation significantly.  In 75\% of cases both $\a0$ and $T$ are determined more accurately by the pq-model (there is no dependence on $T$ or $\a0$).  Fig.~\ref{test15_apeshifts} shows the shift in the estimated AP from the p-model to the pq-model. Shifts are both positive and negative. However, there is a slight tendency, especially at lower $T$, for the pq-model to give lower estimates of both $\a0$ and $T$ (and in general these are also more accurate). 
So at least in this application, the HRD/q factor tends to lower the AP estimates compared to a pure photometric-based estimation.

Although the pq-model yields a systematic error (\mr) half as large as the one from the p-model, it is still statistically significant at around 10$\sigma$ (compare the values in the last two columns of the Table).  As already mentioned in the context of the p-model results, this may be a consequence of small inaccuracies in the MARCS models.  It may also (or instead) be a result of the choice to use the mean of the one-dimensional marginalized PDFs as the estimated AP values. One could make a constant, global correction for this systematic, which is slightly less than 1\% in $T$ and 0.04\,mag in $\a0$ for the pq-model.

As a baseline comparison, Table~\ref{tab:perf_vf05} also shows the summary performance of the \ilium\ algorithm, introduced by Bailer-Jones~\citep{cbj10a}.  \ilium\ performs an iterative, local, nonlinear interpolation of the d-grid via the forward model to estimate the parameters. It essentially tries to locate the peak of a likelihood function. Compared to the p-model, we see that \ilium\ has a similar performance on $T$ but is considerably worse in $\a0$.  (The \ilium\ confidence intervals have been estimated using the covariance transformation -- equation 11 in Bailer-Jones~\citealp{cbj10a}.)  As explained in the \ilium\ paper, the algorithm does not explicitly take into account the degeneracy between the parameters, so this might be a reason for its inferior performance.


\subsection{Interpretation of the HRD/q factor}

We can understand what the HRD/q factor is doing by considering trial solutions in the HRD in Fig.~\ref{hrdmap_vband_comb_z019_8_joinB_forpaper}.  Consider a star with $q=0$ and $\sigma_q=0.1$. If extinction were zero, this would constrain $\mv$ to be $5 \pm 0.1$ and the HRD prior then constrains $T$ to have a distribution given by the corresponding horizontal slice through the HRD.  The likelihood model, for its part, constrains $\a0$ and $T$ from the SED.
At zero extinction this also gives some probability distribution over $T$. The product of this with the first $T$ distribution gives the combined probability distribution over $T$. It is quite possible that this is zero (the two $T$ distributions don't overlap) in which case there is no permitted solution at zero extinction.  If we consider a solution at a higher extinction, then $\mv$ is constrained to a different range by the HRD prior and we get a different distribution over $T$, also from the likelihood model.  
These two one-dimensional distributions over $T$ are just the ``HRD/q factor'' and ``likelihood'' terms in equation~\ref{eqn:pdfat} with fixed $\a0$. The entire two-dimensional PDF is constructed by considering all $\a0$. 

The posterior PDF in equation~\ref{eqn:pdfat} can also be calculated when we have no parallaxes yet still want to use the HRD to constrain solutions. In that case we simply set the $q$ constraint to a constant and recalculate the integral. For want of a name let us call this the {\em ph-model}. The results of applying this to the extended catalogue are also shown in Table~\ref{tab:perf_vf05}. Not surprisingly, its performance lies between that of the p-model and the pq-model. Yet it is encouraging that the HRD prior alone still brings improvement over use of just the colours, by about 13\% in both parameters. This is relevant for classification projects which have no distance information.

\subsection{Impact of metallicity}\label{sect:metallicity}

The position of the stellar locus in the HRD is quite sensitive to the stellar metallicity, a lower metallicity shifting the main sequence down/left in the HRD (to higher effective temperatures) as can be seen in Fig.~\ref{hrd_vband_lowmass_z019_z_0019_forpaper}.  For the experiments discussed so far, adopting solar metallicity for the HRD prior is reasonable, because this matches the atmospheric metallicities inferred by VF05.  It is nonetheless instructive to investigate the impact of adopting a different metallicity.  I reran the pq-model using a prior built with stars of 10-times lower metallicity ($Z=0.0019$, or \feh\,=\,$-$1). This gives systematically higher estimates of the parameters compared to the original pq-model. These shifts are systematic errors with respect to the true parameter values. Quantitatively the shifts are \mr\,=\,+0.03\,dex in $\log T$ (or 400\,K in $T$) and \mr\,=\,+0.35\,mag in $\a0$.  (The precisions are now also worse -- larger \CI\ -- so at least the model recognises there is some kind of problem.)  These systematic errors degrade the overall performance to be worse than the p-model (which does not use the HRD at all). It may be obvious, but introducing additional information into an inference will degrade performance if that information is erroneous.

If we really had no prior estimate of the metallicity (and cannot estimate it from the photometry), then we should play conservative and build an HRD prior covering a range of metallicities. I have done this by combining the same stellar evolution data at the four metallicities $Z=(0.0001, 0.0019, 0.019, 0.3)$ with relative weights of $(0.1, 1, 1, 0.1)$, used in order to down weight the highest and lowest metallicities. To accommodate the large gaps between the loci (Fig.~\ref{hrd_vband_lowmass_z019_z_0019_forpaper}) I simply adopted an 8-times larger kernel bandwidth for the smoothing than was used to create the original HRD prior in Fig.~\ref{hrdmap_vband_comb_z019_8_joinB_forpaper} (there are better ways to do this in a real application). The resulting HRD prior is very broad. It is therefore not surprising that the results of the pq-model based on this are very similar to the p-model results. Accuracy and precision are only improved by a few percent over the p-model. This is just the behaviour we expect from a prior: When the extra information it adds is weak compared to the data, it hardly alters the inference.

To use this HRD prior we should ideally have some estimate of the metallicity which is significantly better than 1\,dex. 
This is virtually impossible with $BVJHK$ photometry when we cannot restrict $T$ and $\a0$ a priori. In contrast, when $\a0$ is known to be less than 1\,mag, it is possible to estimate \feh\ 
from $griz$ photometry to a (mean absolute) accuracy of 0.4\,dex 
for stars with $T$ between 4000 and 10\,000\,K, although at lower metallicities there are even larger systematic errors due to the lack of metallicity signature (C.\ Liu, private communication).
For Gaia it should be possible to estimate metallicity to sufficient accuracy from the low resolution spectroscopy for the brighter hundred million stars: Bailer-Jones~\citep{cbj10a} has shown that even for a broad prior extinction range (0--10\,mag), \feh\ can be estimated to an accuracy of 0.5\,dex at G=15\,mag.
In situations where individual metallicity estimates are not possible, we should make use of any prior information about the population.

\section{Application to 85\,000 Hipparcos/2MASS stars}\label{sect:hip2mass}

In the previous section I constructed a forward model based on real data extended to show variance in $\a0$. I will now use the same model, as well as the same HRD prior and $q$ constraint model, to calculate posterior PDFs for Hipparcos stars which have parallaxes and $BVJHK$ measurements but unknown $A$ and $T$ values. 

In order to avoid extrapolating the forward model too far I only calculate the posterior PDF over the parameter range of the d-grid (4300--7000\,K, 0.0--3.5\,mag).  As the Hipparcos sample is not a priori limited to this range -- we would in particular expect it to include hotter stars -- we cannot expect the model to give good results on stars which lie beyond it.  Some such stars will have photometry outside the range covered by the d-grid and so will yield zero values of the likelihood over the whole d-grid. I refer to these as {\em invalid stars}.  Of the remaining stars, some may give non-zero values of the likelihood, typically at the edge of the d-grid, but could still have true parameters beyond the bounds of the d-grid. I shall refer to these as {\em exterior stars}.  We could attempt to identify and remove these stars based on the input data, for example with a colour cut.  However, this would not take into account the $q$ data, may not remove all exterior stars (or may remove valid stars), and could anyway not be used efficiently with higher dimensional SED data because of the more complex mapping between data and parameters.  I therefore implement a simple way of identifying exterior stars by identifying when their marginalized PDF is strongly truncated at the edge of the grid (explained below).

\subsection{Catalogue cross-matching and data selection (``h2m catalogue'')}

I take parallaxes from the new Hipparcos catalogue of van Leeuwen~\citep{vl07} \citep{vl07b}, which has 117\,955 entries. Using these positions and proper motions, I cross-matched these stars using the Gator tool with the 2MASS catalogue (with a 2\arcsec\ search radius) to obtain the $JHK$ photometry. Finally I used the Hipparcos identification numbers to extract from Simbad the $BV$ data and (for inspection purposes only) the spectral types (where available). 
I retain only those stars which have: complete photometric and astrometric data; best photometric quality in 2MASS (especially SNR\,$>$\,10, {\tt ph\_qual}\,=\,`AAA', {\tt cc\_flg}\,=\,`000'); $\varpi \geq 0.1$\,mas. This results in 
a catalogue of 85\,608 stars which I shall refer to as the {\em h2m catalogue} (Hipparcos/2MASS catalogue).
The 90\% quantile ranges are 6.8--10.2\,mag in $V$, 1--21\,mas in $\varpi$, and $-$6.4 to 0.4\,mag in $q$. 
The uncertainties in $JHK$ are taken from the 2MASS catalogue (most lie between 0.02 and 0.05\,mag). 
I assume 0.02\,mag for the uncertainties in the $B$ and $V$ bands. The correlations in the colours formed from these are again taken into account in the covariance matrix in the likelihood (see Appendix).

\subsection{Results of the p-model and pq-model}

I first applied the p-model to the full h2m catalogue. The posterior PDFs exhibit the same $\a0, T$ degeneracy we saw with the previous data set. Applying the pq-model, we again see that the HRD/q factor reduces the size of the posterior PDF regions, i.e.\ increases the precision of the estimates, although by a much smaller degree than with the previous data: The mean 90\% confidence interval, \CI, reduces from 0.032\,dex to 0.026\,dex for $\log T$, and from 0.36\,mag to 0.29\,mag for $\a0$.
Of the 85\,608 stars in the h2m catalogue, 7406 (9\%)  yielded no solution in the either model because they gave zero likelihood at every point in the d-grid (the {\em invalid stars}).

\begin{figure}
\begin{center}
\includegraphics[width=0.45\textwidth]{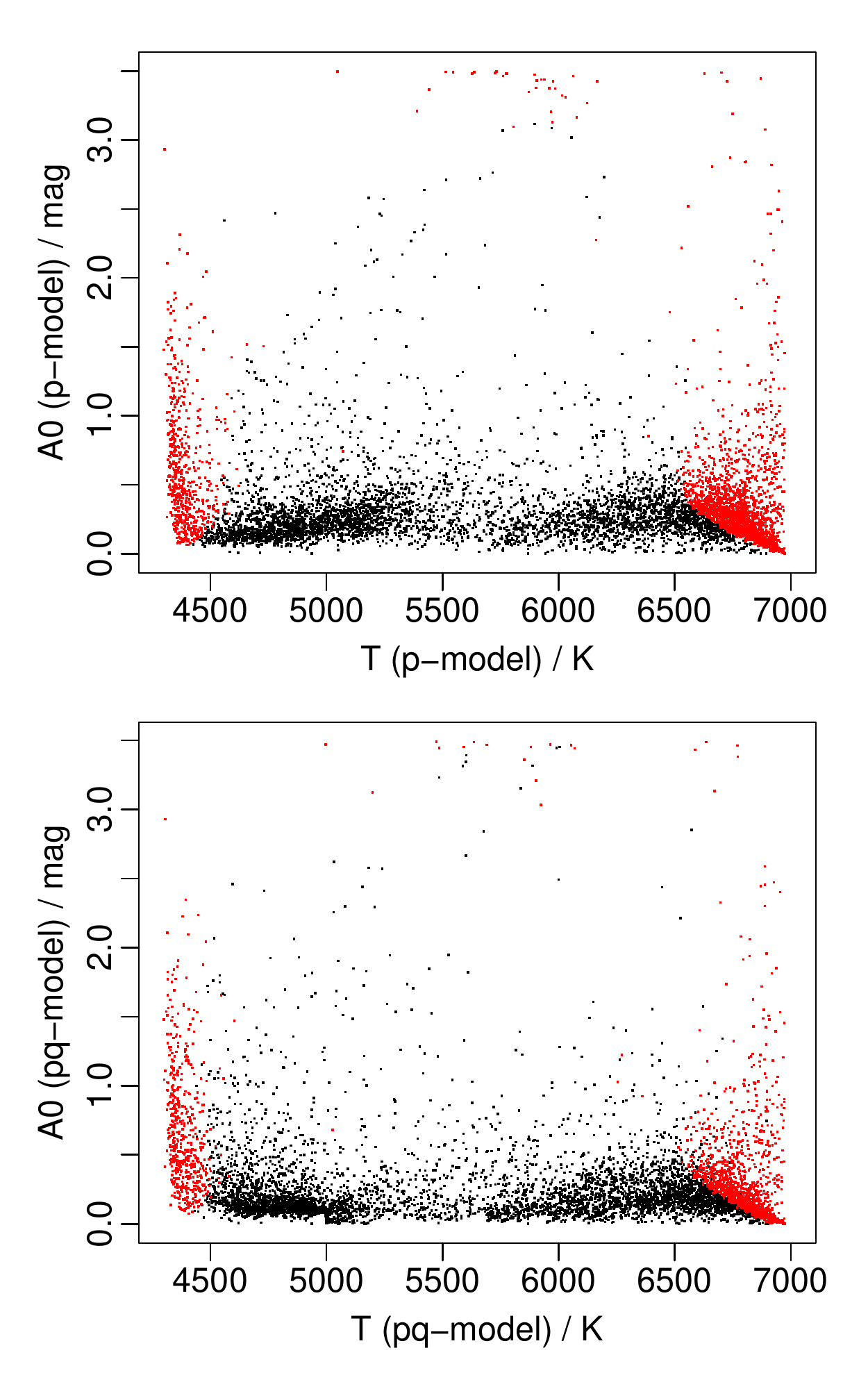}
\caption[]{Estimated mean parameters for the stars in the h2m catalogue inferred from the p-model (top) and pq-model (bottom).
To avoid crowding, these are results based on a randomly selected 12\% of the h2m catalogue, but the general features are present in the full sample. The stars plotted in red are the exterior stars (those with likely true parameters out-of-bounds of the model).
\label{fig:set11_ape}} 
\end{center}
\end{figure}

As before I adopt as the single best parameter estimates for $\a0$ and $T$ the means of the one-dimensional marginalized PDFs. 
Fig.~\ref{fig:set11_ape} shows the distribution of these estimates for the two models.  The vast majority of stars are assigned low extinctions by both models. We also see a paucity of stars at intermediate temperatures.  As expected, there are many solutions lying at the edge of the plot, which are the d-grid boundaries. These stars generally have marginalized PDFs which peak at the edge of the grid and so are effectively truncated by it (similar to the situation in the right-hand column of Fig.~\ref{test15_pdfat_teff_contour_pmodel}).  These probably have true APs which lie outside the grid and therefore beyond the range of applicability of these models.  I identify such a star as an {\em exterior star} if (for either parameter) (a) the peak of the PDF is at the edge of the d-grid, or if (b) the mean of the PDF lies within half of the 90\% confidence interval of the edge of the d-grid. As $\a0$ must be strictly positive, this condition is not applied to the $\a0$\,=\,0\,mag edge of the grid. In this way 39\,678 and 31\,542 exterior stars are identified in the p-model and pq-model results respectively, which is 51\% and 40\% respectively of the valid stars. 
These are overplotted as red points in Fig.~\ref{fig:set11_ape}.  These stars are removed from the subsequent analysis and from the final catalogue provided, leaving 38\,524 stars from the p-model and 46\,660 from the pq-model.

\begin{figure}
\begin{center}
\includegraphics[width=0.45\textwidth]{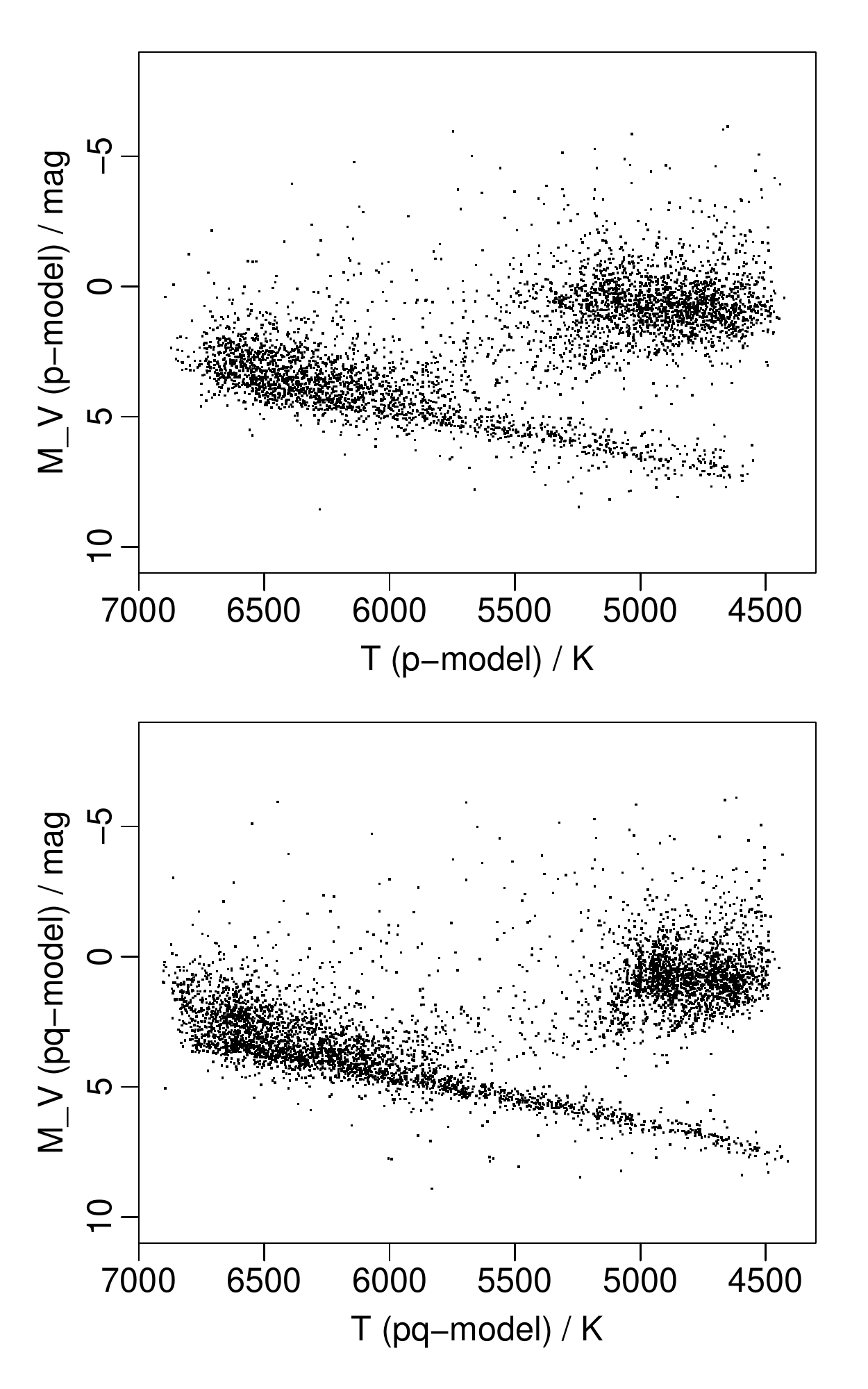}
\caption[]{HRD for the h2m catalogue derived from the p-model (top) and pq-model (bottom), excluding the exterior stars.  To avoid crowding, these are results based on the randomly selected 12\% of the h2m catalogue.
\label{fig:set11_hrd}}
\end{center}
\end{figure}

\begin{figure}
\begin{center}
\includegraphics[width=0.45\textwidth]{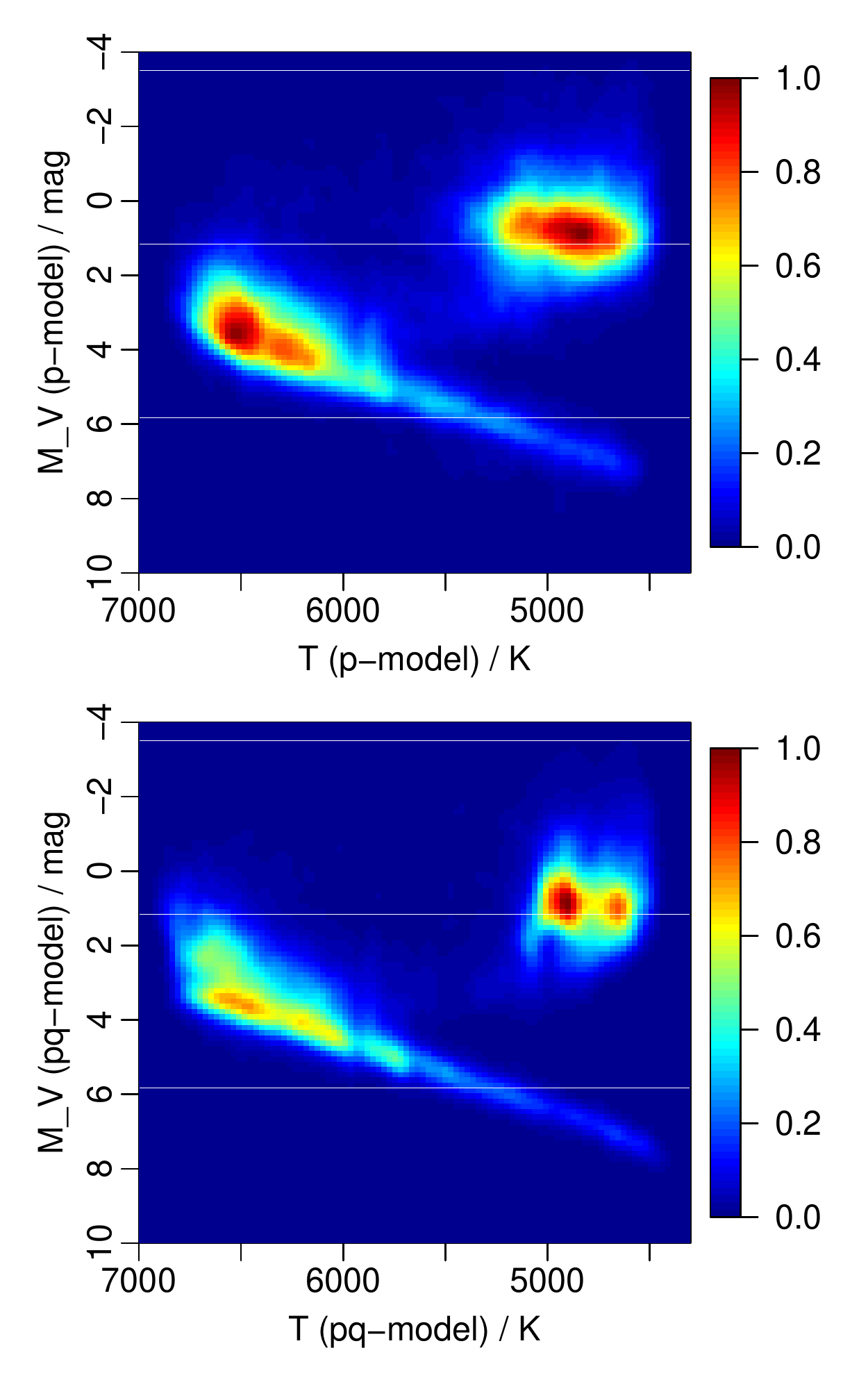}
\caption[]{HRD for the h2m catalogue (excluding exterior stars) derived from the p-model (top) and pq-model (bottom) shown as a density plot (achieved via smoothing with a Gaussian kernel). The number of stars per unit area is normalized to a value of 1.0 at the maximum density (separate normalization in each plot).
\label{fig:set12_hrd_noexterior_density}}
\end{center}
\end{figure}

As we do not have ``true'' parameter estimates for the h2m catalogue I cannot report accuracies for the estimates, so we must assess the results in a different way.  Given the mean parameter estimates I use equation~\ref{eqn:ava0conv} to calculate $\av$ and then calculate the absolute magnitude for each star (from equation~\ref{eqn:ma_constraint2}).\footnote{A passionate Bayesian would propagate PDFs to the end, marginalize over $\a0$ and $T$ and then calculate the mean of $P(\mv | \vecp, q)$.}  The resulting HRD for the h2m stars is shown in Fig.~\ref{fig:set11_hrd} for parameters from the p-model (upper panel) and pq-model (lower panel).  This plot does not reflect the relative densities in the HRD very well, so it is redrawn (using the full sample) as a density scale plot in Fig.~\ref{fig:set12_hrd_noexterior_density} (note that a smaller $\mv$ scale is used).
Both HRDs show a clear separation between a main sequence, dominated by stars between 5700\,K and 6700\,K, and a giant branch at around 4500--5000\,K. (We likewise see two distinct populations with different colours in colour--colour diagrams.) The HRDs are quite similar overall, but there are important differences. The main sequence in the pq-model has a narrower $\mv$ spread for stars cooler than solar, but a broader spread for hotter stars. The giant branch in the pq-model has a narrower temperature spread and also displays two density maxima (discussed in the next section).
An order of magnitude check on the correctness of these HRDs can be made by considering the position of typical stars.
The Sun, for example, has $T$\,=\,5800\,K and $\mv$\,=\,4.8\,mag, which would correctly place it on the main sequence in both models.

If we now apply the ph-model to these data (that is, we use the HRD prior but not the $q$ factor), then the resulting HRD density map is actually quite similar to that arising from the p-model. This proves that most of the difference in the HRD morphology we see in Fig.~\ref{fig:set12_hrd_noexterior_density} comes from the information provided by the parallax in combination with the HRD, rather than from the HRD alone.

\begin{figure}
\begin{center}
\includegraphics[width=0.45\textwidth]{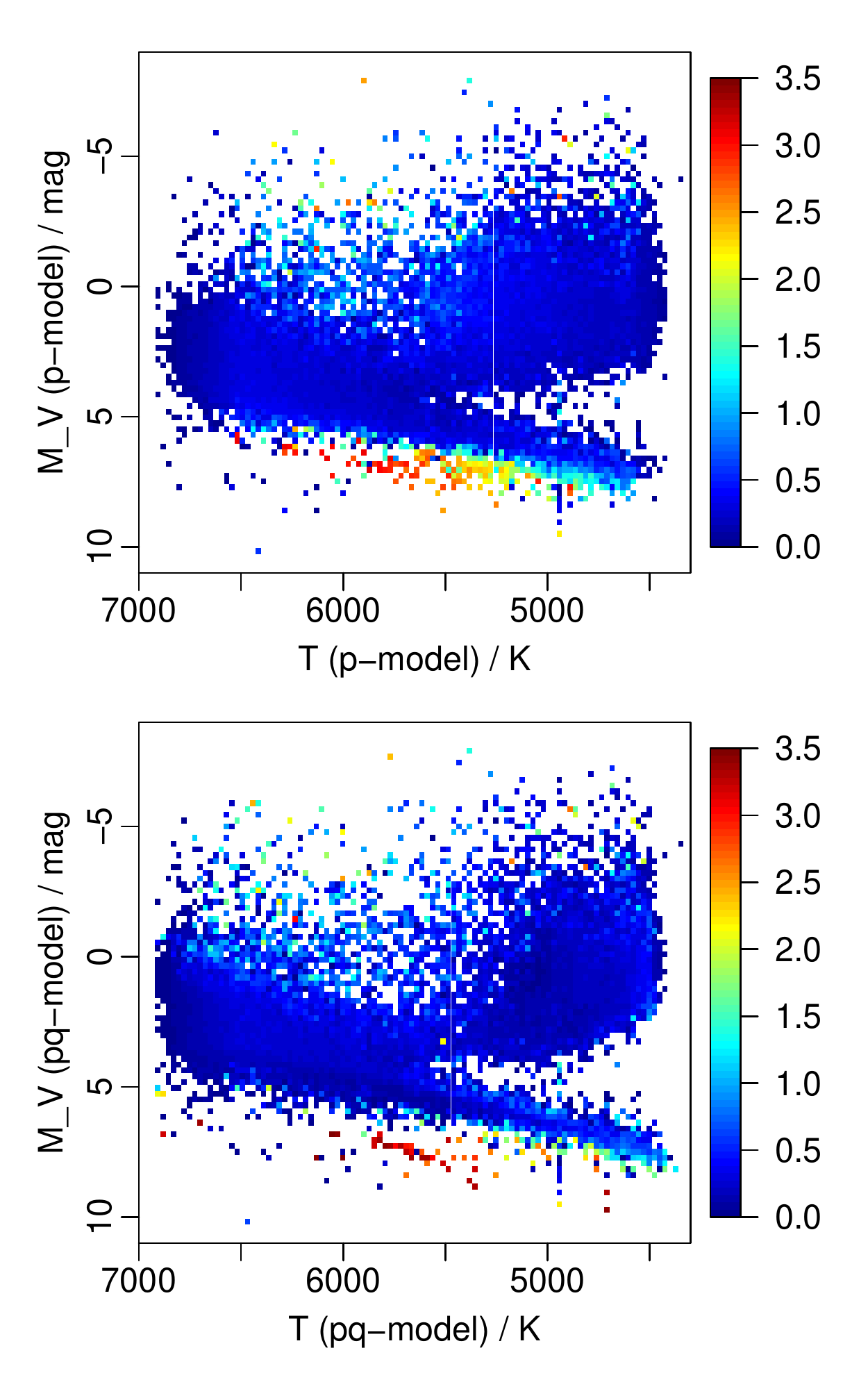}
\caption[]{HRD for the complete h2m catalogue (excluding exterior stars) derived from the p-model (top) and pq-model (bottom), showing the mean extinction ($\a0$ in magnitudes) on a colour scale. Unoccupied areas are shown in white.
\label{fig:set12_hrd_noexterior_a0colscale_fulla0range}}
\end{center}
\end{figure}
 
\begin{figure}
\begin{center}
\includegraphics[width=0.45\textwidth]{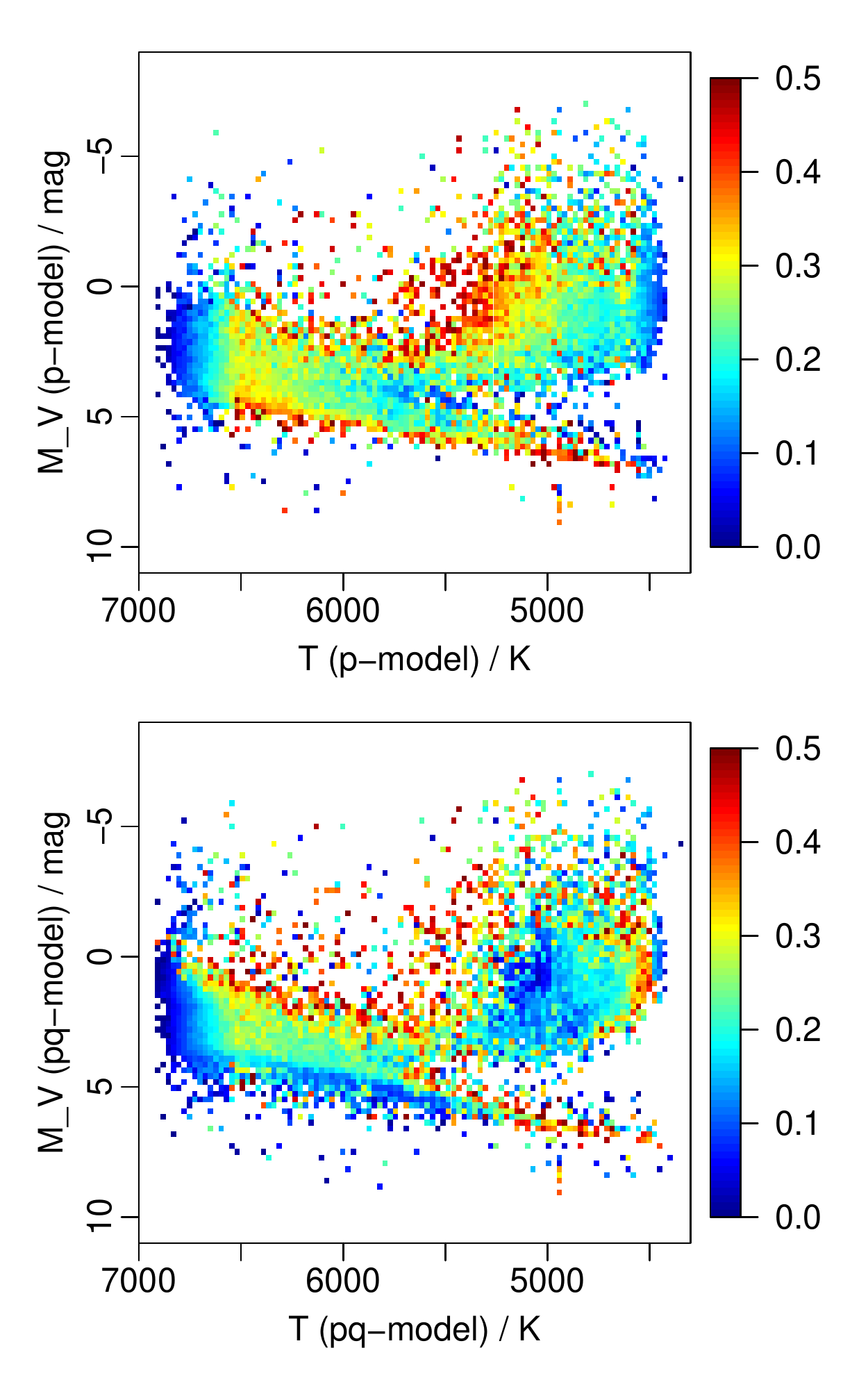}
\caption[]{As Fig.~\ref{fig:set12_hrd_noexterior_a0colscale_fulla0range} but showing a narrower range of low extinctions (0-0.5\,mag). Cells of higher mean extinction are not plotted (about 10\% of stars in both models).
\label{fig:set12_hrd_noexterior_a0colscale}}
\end{center}
\end{figure}
 
Fig.~\ref{fig:set12_hrd_noexterior_a0colscale_fulla0range} which shows the HRD using a colour scale to represent the mean extinction at each point in the diagram. We see only a few regions with high extinctions, predominantly stars below the main sequence. 
Fig.~\ref{fig:set12_hrd_noexterior_a0colscale} shows the same but increasing the dynamic range for the low extinction stars (0--0.5\,mag), which are 90\% of stars in both models. 

\subsection{Discussion} 


\begin{figure}
\begin{center}
\includegraphics[width=0.45\textwidth]{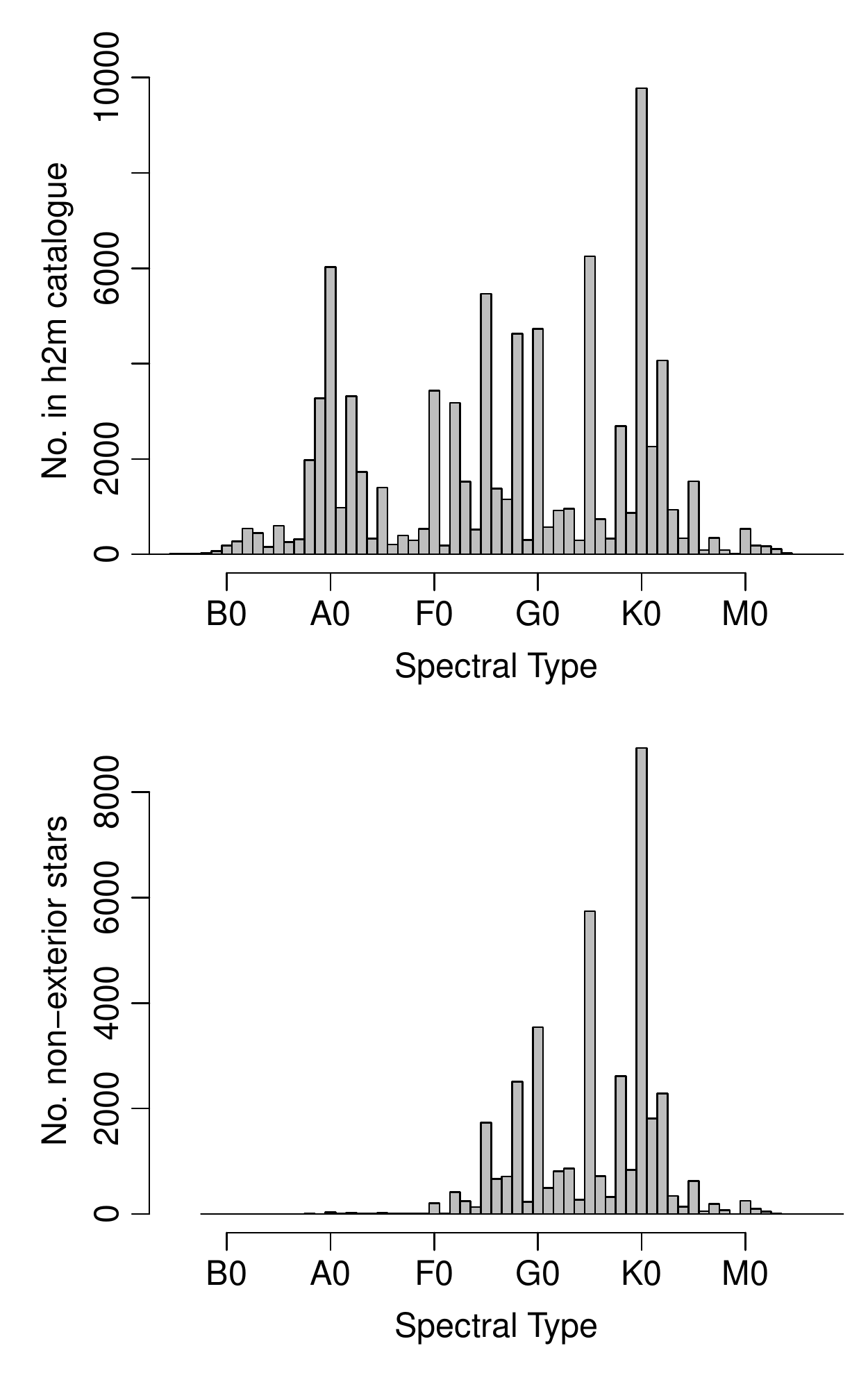}
\caption[]{Distribution of Simbad spectral types for all stars in the h2m catalogue (top) and for the
valid, non-exterior stars as identified by the pq-model (bottom).
\label{fig:set12_spthist}} 
\end{center}
\end{figure}
 
\begin{figure}
\begin{center}
\includegraphics[width=0.45\textwidth]{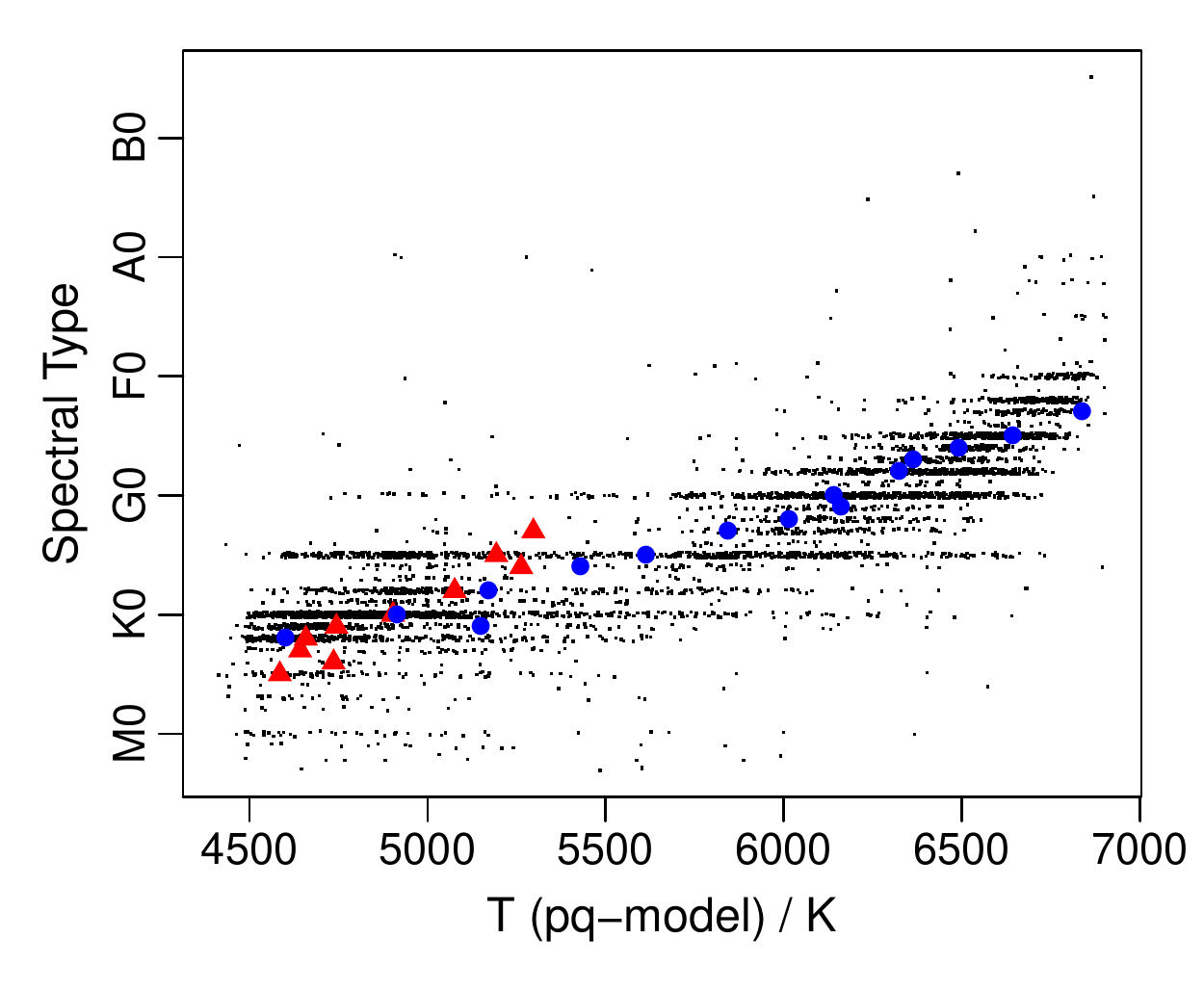}
\caption[]{Relationship between Simbad spectral types and effective temperature inferred by the pq-model for the h2m catalogue. To avoid crowding, results for only the randomly selected 12\% of the catalogue are show, plus points have been jittered by $\pm \frac{1}{4}$ spectral types. Blue circles and red triangles show the calibration for dwarfs and giants (respectively) from Bailer-Jones et al.~\citep{cbj97}.
\label{fig:set11_ape_spt}} 
\end{center}
\end{figure}
 
Simbad provides spectral types for the majority of the h2m stars. These inevitably come from heterogeneous sources, yet provide a rough check on the inferred temperatures. The spectral types are provided as character strings, sometimes with uncertain or intermediate types such as {\tt F6/F7V}, in which case I just take the first spectral type. The distribution of spectral types in the whole h2m catalogue is shown in the upper panel Fig.~\ref{fig:set12_spthist}. Removing the invalid and exterior stars for the pq-model leaves those plotted in the lower panel. As expected this has removed predominantly early-type (hot) stars. Fig.~\ref{fig:set11_ape_spt} shows the relationship between these spectral types and the inferred $T$ estimates for this model. The correlation is reasonable, especially when we consider that there is anyway no tight relationship between $T$ and spectral type a priori.
As comparison, this figure also shows the average $T$--spectral type calibration for dwarfs and giants from 
Bailer-Jones et al.~\citep{cbj97} for the Galactic thin disk at \feh\,=\,$-$0.2\,dex (there is a scatter of around $\pm$300\,K at each spectral type in the calibration). The agreement is reasonable, supporting the conclusion that the $T$ assignments from the pq-model are appropriate.

Simbad also supplies a luminosity class for about half of the h2m stars. We can use these to make a check on the apparent clear separation between a main sequence (dwarf stars) and giant branch (giant stars) in Fig.~\ref{fig:set12_hrd_noexterior_density}.  For this purpose I make a simple classification of the stars as giants if both $T<5250$\,K and $\mv < 4$\,mag, and as dwarfs otherwise. For the pq-model this classifies 39\% of the stars as giants.  Of these, 96\% of those which have a luminosity class are class {\tt III}, {\tt II/III} or {\tt III/IV}, i.e.\ a giant class.  Of the dwarfs, 93\% of those which have a luminosity class have are class {\tt V} or {\tt IV/V}, i.e.\ a dwarf class.  
Assuming we have roughly correct temperature assignments, this suggests that the extinction estimates cannot be so wrong as to have mixed up dwarfs and giants in the cooler part of the HRD.


The inferred extinction distribution of the stars shown in Fig.~\ref{fig:set11_ape} is consistent with the three-dimensional extinction model 
of Vergeley et al.~\cite{vergely97}. This model, which is 
based on Hipparcos parallaxes with extinctions estimated from Str\"omgren photometry, gives an average extinction ($\av$) in the Galactic plane in the solar neighbourhood of 1.5 mag/kpc (with a vertical scale height of 70\,pc). This is compatible with the present result: most of the h2m stars have extinctions below 0.5\,mag, and most lie between a distance of 50\,pc and 1\,kpc.


In both panels of Fig.~\ref{fig:set12_hrd_noexterior_a0colscale_fulla0range} we see some highly extinct stars below the main sequence. 
In the pq-model there are 41 such stars with $\mv>5$\,mag and $\a0>3.0$\,mag.  They are all very nearby (4--18\,pc) and faint ($V$\,=\,9.6--12.0\,mag) and have very red $V\!-\!J$ colours: 3.4--4.2\,mag for 37 of them, whereas only 0.2\% of the valid, non-exterior stars in the pq-model have such red colours. This is at the limit of the colours used to fit the forward model (Fig.~\ref{v2ext_col_vs_teff}). 
These stars also have very late spectral types of M0--M4.5 from Simbad. Assuming these spectral types to be correct, this would correspond to effective temperatures between about 3800 and 3000\,K (Leggett et al.~\citealp{leggett00}). This is well below the lower limit of the extended catalogue (4707\,K), so the inference would depend on a significant extrapolation of the forward model. 
As these colours and APs are not represented in the model training we cannot expect a reliable inference, so I conclude that the pq-model and p-model AP assignments are incorrect. Ideally these stars would have been identified as ``exterior'' stars, yet it is unsurprising that some slipped through the somewhat ad hoc criteria adopted.


We saw in section~\ref{sect:metallicity} that adopting an HRD prior with a lower
metallicity than the true metallicity resulted in a systematic increase (error) in both $\a0$ and $T$.  A priori we expect some of the Hipparcos stars to have subsolar metallicity, in which case the use of a solar metallicity HRD prior for these stars is expected to yield the converse, that is, erroneously low $\a0$ and $T$ estimates. 
A lower extinction corresponds to a larger intrinsic magnitude on account of the $q$ constraint, although for the ten times lower metallicity used in section~\ref{sect:metallicity} the change in $\a0$ and therefore in $\mv$ was only 0.3\,mag on average.

\begin{figure}
\begin{center}
\includegraphics[width=0.438\textwidth]{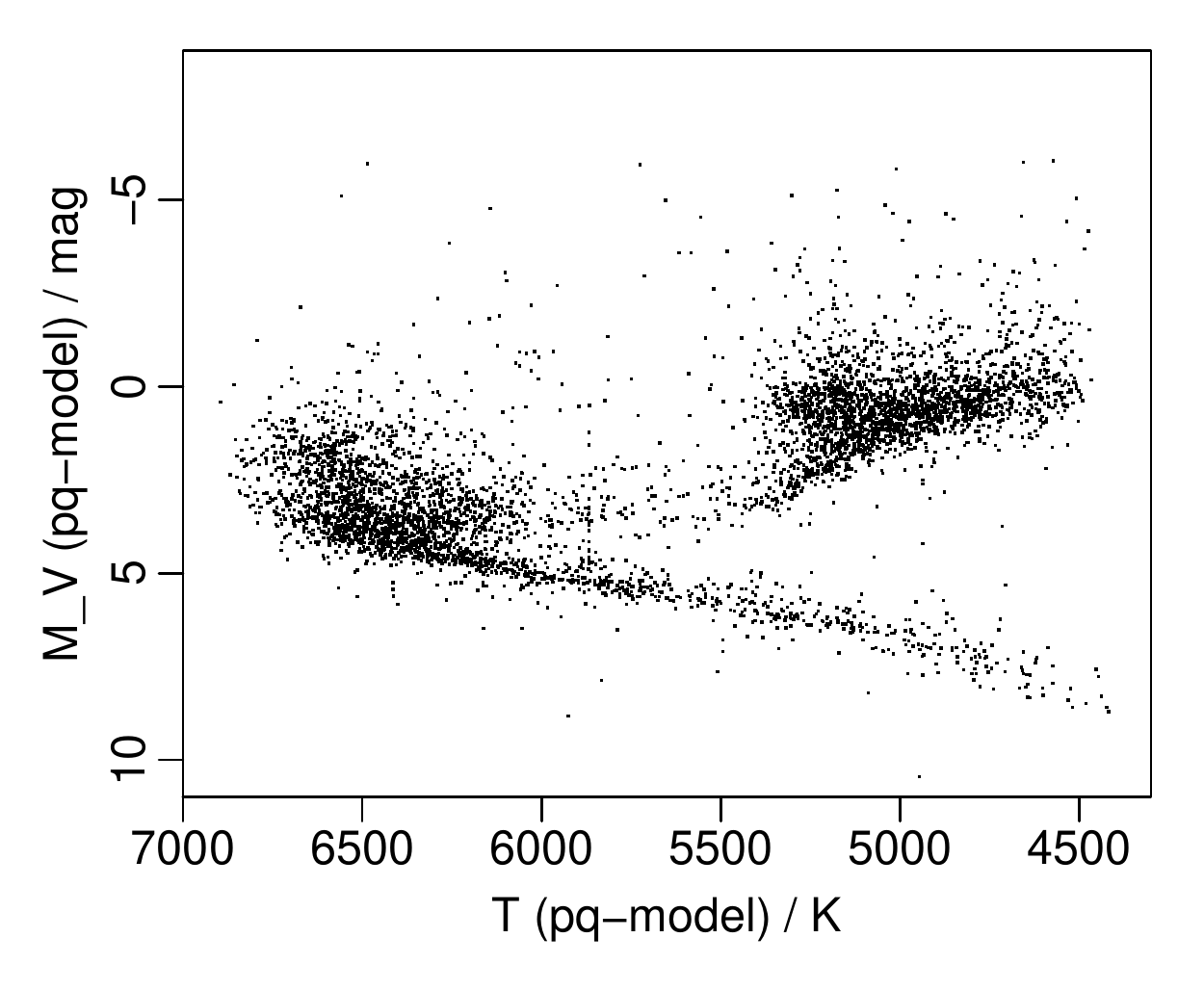}
\caption[]{HRD for the h2m catalogue derived from the pq-model using the HRD prior with $Z=0.0019$, i.e.\ ten times below solar metallicity. Exterior stars are excluded. 
\label{fig:set13_hrd}}
\end{center}
\end{figure}

I investigated the impact of metallicity empirically by rerunning the pq-model using the ten times lower metallicity prior. The resulting HRD is shown in Fig.~\ref{fig:set13_hrd} and should be compared with the lower panel of Fig.~\ref{fig:set11_hrd}. 
We do indeed see that the main sequence has been shifted to a slightly larger $\mv$. The main sequence is also not as tight as before and the upper part of the main sequence has a different distribution. There is also a significant change in the morphology of the giant branch. In particular it now extends to higher temperatures.


We saw in Fig.~\ref{fig:set12_hrd_noexterior_density} that the pq-model HRD exhibits a bimodality in the giant branch. This might reflect two genuine populations with different metallicities, as was inferred for the Hipparcos sample by Girardi et al.~\citep{girardi98}. \footnote{The bimodality is not seen in the p-model HRD, perhaps because the APs are less precise (and presumably less accurate) in that model, but perhaps because the HRD/q factor is providing a better discrimination between the two populations.}
Assuming one population has near-solar metallicity and the other sub-solar metallicity, the latter would be
the bluer (higher $T$) of the two. However, as just discussed, the use of the solar metallicity HRD prior results in such stars being
assigned erroneously low $\a0$ and $T$ values. (So if this higher $T$ clump really is low metallicity, in reality it should have higher $T$, higher $\a0$ and lower $\mv$.) We see from Fig.~\ref{fig:set12_hrd_noexterior_a0colscale} that the higher $T$ clump is indeed assigned a lower $\a0$. Assuming that, in reality, there is no systematic change in the extinction across the giant branch, this apparently lower extinction may in fact be a result of an extinction--metallicity confusion (degeneracy) introduced by using an erroneous metallicity in the HRD prior in combination with the $\a0$--$T$ degeneracy. This again underlines the importance of estimating metallicity from the SED.

\subsection{Catalogue} 

\begin{table*}
\setlength{\tabcolsep}{3pt} 
\begin{center}
\begin{minipage}{0.9\textwidth}
\caption{A sample of the output catalogue (rows 5000--5010). The columns are defined in Table~\ref{catcol}\label{pubcat}. NA stands for ``not available'', i.e.\ it is an exterior star in that model. The full version is available online, e.g. from http://www.mpia.de/homes/calj/qmethod.html}
{\scriptsize
\begin{tabular}{rrrrrrrrrrrrrrrrrrrrrrr}
\hline
1 & 2 & 3 & 4 & 5 & 6 & 7 & 8 & 9 & 10 & 11 & 12 & 13 & 14 & 15 & 16 & 17 & 18 & 19 & 20 & 21 & 22 & 23 \\
\hline
10882 & 3.7291 & 3.6982 & 3.7607 & 0.39 & 0.00 & 0.77 & 3.6998 & 3.6938 & 3.7040 & 0.02 & 0.00 & 0.10 & 9.71 & 8.86 & 7.178 & 0.021 & 6.738 & 0.029 & 6.588 & 0.017 &  2.67 & 0.80 \\
10883 &     NA &     NA &     NA &   NA &   NA &   NA & 3.7865 & 3.7591 & 3.8107 & 0.34 & 0.03 & 0.61 &10.69 &10.14 & 8.866 & 0.023 & 8.519 & 0.042 & 8.460 & 0.021 &  9.56 & 1.28 \\
10886 & 3.7540 & 3.7313 & 3.7801 & 0.23 & 0.00 & 0.52 & 3.7588 & 3.7334 & 3.7835 & 0.28 & 0.00 & 0.57 & 8.84 & 8.10 & 6.736 & 0.025 & 6.381 & 0.027 & 6.261 & 0.016 & 12.61 & 0.89 \\
10887 & 3.6790 & 3.6606 & 3.7008 & 0.19 & 0.00 & 0.48 & 3.6739 & 3.6597 & 3.6943 & 0.12 & 0.00 & 0.41 & 9.61 & 8.51 & 6.601 & 0.021 & 6.079 & 0.026 & 5.933 & 0.022 &  2.47 & 1.14 \\
10888 & 3.8210 & 3.7980 & 3.8404 & 0.24 & 0.00 & 0.46 & 3.8181 & 3.7963 & 3.8380 & 0.20 & 0.00 & 0.43 & 9.48 & 9.02 & 8.056 & 0.043 & 7.871 & 0.031 & 7.815 & 0.020 &  6.31 & 1.19 \\
10890 &     NA &     NA &     NA &   NA &   NA &   NA & 3.8227 & 3.7988 & 3.8385 & 0.27 & 0.01 & 0.45 & 7.09 & 6.65 & 5.665 & 0.018 & 5.452 & 0.026 & 5.440 & 0.021 & 15.93 & 0.56 \\
10891 & 3.7094 & 3.6794 & 3.7360 & 0.40 & 0.01 & 0.74 & 3.6929 & 3.6759 & 3.7036 & 0.18 & 0.00 & 0.35 &10.43 & 9.48 & 7.630 & 0.020 & 7.119 & 0.026 & 7.021 & 0.021 &  1.88 & 0.98 \\
10894 & 3.7397 & 3.7051 & 3.7721 & 0.43 & 0.01 & 0.82 & 3.7456 & 3.7045 & 3.7780 & 0.50 & 0.01 & 0.88 &10.40 & 9.56 & 7.939 & 0.027 & 7.511 & 0.055 & 7.370 & 0.027 &  2.63 & 1.48 \\
10895 & 3.6650 & 3.6395 & 3.6853 & 0.35 & 0.01 & 0.62 & 3.6661 & 3.6452 & 3.6841 & 0.36 & 0.10 & 0.61 & 9.23 & 8.00 & 5.863 & 0.017 & 5.290 & 0.016 & 5.159 & 0.018 &  3.81 & 0.96 \\
10899 & 3.6746 & 3.6596 & 3.6881 & 0.16 & 0.00 & 0.35 & 3.6706 & 3.6592 & 3.6852 & 0.10 & 0.00 & 0.31 & 8.84 & 7.72 & 5.804 & 0.018 & 5.275 & 0.018 & 5.118 & 0.017 &  4.12 & 0.48 \\
10900 & 3.6690 & 3.6556 & 3.6823 & 0.11 & 0.00 & 0.29 & 3.6672 & 3.6562 & 3.6770 & 0.09 & 0.00 & 0.23 & 9.12 & 7.99 & 6.079 & 0.020 & 5.516 & 0.024 & 5.369 & 0.020 &  3.79 & 0.89 \\
\hline
\end{tabular}
}
\end{minipage}
\end{center}
\end{table*}

\begin{table}
\begin{center}
\caption{Definition of the columns in the output catalogue (CI stands for ``confidence interval'')\label{catcol}}
\begin{tabular}{ll}
\hline
1 & Hipparcos identifier \\
2 & $\log T$/K from the p-model (mean estimate) \\
3 & lower bound of the 90\% CI to $\log T$/K from the p-model \\
4 & upper bound of the 90\% CI to $\log T$/K from the p-model \\
5 & $\a0$/mag from the p-model (mean estimate) \\
6 & lower bound of the 90\% CI to $\a0$/mag from the p-model \\
7 & upper bound of the 90\% CI to $\a0$/mag from the p-model \\
8 & $\log T$/K from the pq-model (mean estimate) \\
9 & lower bound of the 90\% CI to $\log T$/K from the pq-model \\
10 & upper bound of the 90\% CI to $\log T$/K from the pq-model \\
11 & $\a0$/mag from the pq-model (mean estimate) \\
12 & lower bound of the 90\% CI to $\a0$/mag from the pq-model \\
13 & upper bound of the 90\% CI to $\a0$/mag from the pq-model \\
14 & $B$ magnitude \\
15 & $V$ magnitude \\
16 & $J$ magnitude \\
17 & $1\sigma$ uncertainty in the $J$ magnitude \\
18 & $H$ magnitude \\
19 & $1\sigma$ uncertainty in the $H$ magnitude \\
20 & $K$ magnitude \\
21 & $1\sigma$ uncertainty in the $K$ magnitude \\
22 & parallax / mas \\
23 & $1\sigma$ uncertainty in the parallax / mas \\
\hline
\end{tabular}
\end{center}
\end{table}

A catalogue of the AP estimates for 46\,900 stars are available online with the electronic version of this article and from CDS Strasbourg.  38\,524 stars have APs from the p-model and 46\,660 have APs from the pq-model (most have APs from both).  $T$ spans 4350--6900\,K and $\a0$ 0--3.45\,mag.  A sample of the catalogue is show in Table~\ref{pubcat} with the columns defined in Table~\ref{catcol}. Columns 2--13 are the estimates from the present work. {\tt NA} is used to indicate where either the p-model or pq-model does not provide parameter estimates (i.e.\ it is an {\em exterior star}, as defined in section~\ref{sect:hip2mass}; {\em invalid stars} are excluded from the catalogue).  Columns 1, 22 and 23 are from the catalogue of van Leeuwen~\citep{vl07}, columns 14 and 15 are from Simbad and columns 16--21 are taken from the 2MASS catalogue.

To convert the extinction parameter, $\a0$, to the extinction in the $V$ band, $\av$, one may use the following quadratic approximation to the function $y(\a0, T)$ in equation~\ref{eqn:ava0conv}
\begin{eqnarray}
y(\a0, T) \!\!\! &\simeq& \!\!\! -5.376 + 2.884(\log T) - 0.4217\a0 \\ 
                 && \!\!\!  - 0.3865(\log T)^2 - 0.00374\a0^2 + 0.1072(\log T)\a0 \nonumber \ .
\end{eqnarray}
The root-mean-square error of this fit is 0.0025\,mag over the fitted parameter range 4000--7000\,K and 0--5\,mag in $\a0$.

\section{Assumptions and possible extensions}\label{sect:extensions}

The method as presented could be improved and extended in a number of ways. 

The range over which we can estimate APs is set by the range of validity of the forward model, which in turn depends on the labelled templates used to fit it.  I focused on F,G,K stars. This could be extended using additional data at higher temperatures, although we must be careful to ensure that they are parametrized on a consistent temperature scale.
Preliminary predictions of the method's performance on simulated Gaia spectrophotometry and astrometry for stars over a wider $\a0$ and $T$ range are reported in Bailer-Jones~\citep{cbj10b}.

The development of the method here has assumed for simplicity that stars are described by just three APs: $T$, $\a0$ and $\mv$.  The SED ($\vecp$) is assumed to depend on only the first two of these, with dependence on $\mv$ introduced by the measured apparent magnitude and parallax ($q$).  
I only derived PDFs over $T$ and $\a0$, choosing then to simply derive $\mv$ using equation~\ref{eqn:ma_constraint2}, but we could also derive a PDF over $\mv$.

Given higher resolution data, we can usually estimate additional parameters from the SED, in particular the metallicity, \feh, and the surface gravity, \logg.  In the demonstrations above, the p-model essentially assumes all stars to be solar metallicity dwarfs, 
because this is what was used to fit the forward model. Although this is not correct for the h2m data, it is nonetheless acceptable for the broad band colours considered here because they show little sensitivity to either metallicity or surface gravity. When it comes to the pq-model, however, we have seen that the assumed metallicity, $Z$, of the HRD does have a significant impact on the inferred $T$ and $\a0$, so we should introduce some dependence on $Z$ in the method. If we replace $T$ with $(T, Z)$ in equation~\ref{eqn:a01} and follow the derivation through, we arrive again at equation~\ref{eqn:pdfat} but with an additional $Z$ in the conditioned terms (to the right of the ``$|$'') for the likelihood and the HRD prior. It does not explicitly appear in the $q$ constraint because once conditioned on $\mv$, $Z$ adds no further information. For the likelihood term we would estimate atmospheric \feh\ and then convert to $Z$ assuming some fixed abundance pattern (although in principle we could introduce the Helium abundance, $Y$, as another parameter in the HRD).
Estimating $Z$ to around 0.5\,dex or better is equivalent to using a ``narrower'' HRD prior, which in turn would lead to much more
precise estimates of $T$, $\a0$ and $\mv$.

We can likewise extend the method to estimate the surface gravity, \logg, which is often estimated from higher resolution spectroscopy and would help to constrain $\mv$ further. \logg\ would appear in the same way as $Z$ in equation~\ref{eqn:pdfat}, but unlike $Z$ it is not an independent parameter in the HRD, but rather is determined by $\mv$ and $T$. 


For a large, deep survey like Gaia it may be of particular importance to permit the selective extinction parameter, $\r0$ to vary, as it known to vary across the Galaxy (e.g.\ Patriarchi et al.~\citealp{patriarchi03}). This could easily be introduced via simulation in the same way that $\a0$ variance was introduced in section~\ref{sect:datacon}. An $\r0$ dependence is introduced into equation~\ref{eqn:pdfat} by replacing $\a0$ with $(\a0, \r0)$.  To be effective we need to be able to estimate $\r0$ from the SED. First investigations by the MPIA Gaia group suggest that this will be possible with the Gaia low resolution spectroscopy (C.\ Liu, private communication), although the degeneracies with $\a0$ and $T$ are yet to be fully characterized.

Finally, the method as presented is predicated on all stars being single.  A physical binary comprising two identical stars has $\mv$ which is 0.75\,mag lower than a single star of the same $T$. Such binaries could be accommodated using an adjusted HRD and introducing an extra AP, $s$, the probability that the star is single. To solve for the other APs (the common $T$ and $\a0$) we would marginalize over $s$, adopting the single star HRD when $s\!=\!1$ and an HRD shifted by 0.75\,mag when $s\!=\!0$. We can likewise infer $P(s | {\vecp}, q)$.
The situation is more complex when we allow for more general types of binary, but can be helped if we can make some estimates of the effective temperatures of the two components from their composite spectrum. The MPIA Gaia group has had some limited success in this with simulated Gaia spectra (P.\ Tsalmantza, private communication).

In all of this work, there is choice in the nature of the HRD prior (just as there is a choice in what SED we observe).  The HRD prior was adopted primarily to eliminate the ``forbidden'' regions of the HRD (the white regions in Fig.~\ref{hrdmap_vband_comb_z019_8_joinB_forpaper}), rather than to introduce specific dependence on the IMF or star-formation rate used to construct it.  All pq-model results were for a power-law (Salpeter) IMF, but we actually get very similar results with a flat IMF (which is not that surprising, given the relatively narrow $T$ and hence mass range).
 

It is worth noticing that the HRD prior is a prior on the absolute magnitude and not on the observed magnitude. It 
therefore does not take into account the magnitude limit of the survey from which the data have been obtained. This could be potentially useful information. For example,  the limiting magnitude may be such that we do not expect to find very many intrinsically faint white dwarfs or brown dwarfs. 

The whole approach adopted has been to estimate APs for single stars by considering their line-of-sight extinction as an intrinsic parameter. This ignores any correlation in the extinction between neighbouring stars.  Yet in many parts of the Galaxy it may be reasonable to assume that the density of interstellar material, and hence the line-of-sight extinction to that point, varies smoothly with three-dimensional position.  We could introduce this smoothness constraint through a hierarchical inference procedure in which we solve for the APs of a set of $N$ stars simultaneously. The idea is to parametrize the extinction as a function of (known) spatial position, $\a0(l, b, \varpi ; \vecbeta)$, where $l,b$ are the Galactic coordinates and $\vecbeta$ are the free parameters of this function.  Rather than solving for $N$ two-dimensional PDFs over $(T, \a0)$ for each star independently, we solve for the much higher dimensional PDF over $(T_1 \ldots T_N, \vecbeta)$. This is computationally far more complex, as it involves $N$ non-independent integrals like that in equation~\ref{eqn:pdfat}. But it constrains the extinction to vary in a physically plausible way, plus it would help to estimate extinction in those cases where it may otherwise be very difficult.
Such a large-scale inference is not unthinkable for a survey such as Gaia.

Let us finally consider the method's philosophy of separating the observed data into independent terms $\vecp$ and $q$. This leads to an astrophysically-meaningful and useful separation of the terms in equation~\ref{eqn:pdfat}, in which 
we can simply use the HRD/q factor to ``update'' the AP estimates obtained from the pure SED (equation~\ref{eqn:a07}). 
The price we pay for this is the need to normalize the SED (achieved here by forming colours). This introduces correlations between the elements of the SED, although this was easy to accommodate in the likelihood function in the present case. If we were willing to sacrifice the separability, then we could adopt a more purist approach and operate one step closer to the raw data directly on the fluxes and the parallax.  This would be of little benefit for Gaia, however, because the parallax and the apparent ($G$-band) magnitude are derived from the same measurements so are anyway correlated. Furthermore, the planned processing of the Gaia SED (the ``BP/RP spectrum'') will probably also result in correlated fluxes.

\section{Conclusions}

The main conclusions of this work and features of the method are as follows
\begin{itemize}

\item there exists a significant degeneracy between $\a0$ and $T$ when estimated from broad band optical/infrared data; 

\item the method developed here allows one to make self-consistent and quantitative use of the HRD prior and parallaxes in order to improve the accuracy of the AP estimation beyond using just the colours (by 35\% overall with the extended catalogue);

\item even if parallaxes are not available, use of the HRD prior improves AP accuracy compared to using just the colours (by 13\% overall with the extended catalogue);

\item the method gives a multidimensional posterior density distribution over the APs and can easily be extended to include other APs and applied to other (spectro)photometric data;

\item the method ensures that all derived parameters are self-consistent. It takes into account uncertainties and covariances in the data;

\item accurate estimation of metallicity (to about 0.5\,dex) is necessary if the use of the HRD/q factor is to improve the accuracy of the $T$ and $\a0$ estimation.

\end{itemize}

\noindent
The method developed in this paper has significant advantages over standard pattern recognition methods, such as neural networks or support vectors machines. These try to learn an inverse mapping from the input data to the APs, yet this would be cumbersome to achieve with these heterogeneous photometric and parallax inputs. First, we would have to ensure that the stars in the training set have consistent magnitude, parallax, SED and extinction.  Second, we would have to train separate algorithms depending on what inputs we use (as these methods are not robust to simply omitting input variables). The method developed in this paper, in contrast, has no need for any training data on, or modelling of, the parallax or apparent magnitude (only their uncertainties are modelled).
Pattern recognition methods normally also give just a single solution and are incapable of naturally providing probability distributions (and thus expected uncertainties) over parameters. They likewise cannot easily incorporate prior information. Most signiﬁcantly, however, these methods cannot explicitly take into account the constraints from physical background information we have, namely the relationship between temperature, extinction, parallax and magnitude, which we know from stellar physics and geometry. 
Approaches which do not take these constraints properly into account could produce unphysical solutions.
Physical constraints and other prior information could only be incorporated indirectly (and probably incompletely) via some clumsy tuning of the training data set or transformation of the variables. The only genuine advantage that bare pattern recognition methods offer is speed, but the saving this offers in terms of computer power is hardly significant compared to the cost of gathering astrometric data.

\section*{Acknowledgements}

I am grateful to Antonella Vallenari for providing me with output from her stellar population models for building the HRD prior.  For constructive criticism and comments I would like to thank Jo Bovy, Ron Drimmel, David Hogg, Dustin Lang and Antonella Vallenari.  I am grateful to Chris Stubbs, my host at Harvard, and the other members of the LPPC for supporting my sabbatical stay where a substantial part of this project was conducted. This work makes use of Hipparcos and 2MASS data, and has used the Simbad and Vizier services at CDS, Strasbourg as well as the NASA/IPAC Infrared Science Archive (IRSA) for constructing the data sets.

\appendix

\section{Covariance}

Let ${\rm Var}()$ denote the variance and ${\rm Cov}()$ the covariance operators.
For two random variables $X$ and $Y$ with arbitrary distributions, a general result is
\begin{equation}
{\rm Var}(X+Y) = {\rm Var}(X) + {\rm Var}(Y) + 2{\rm Cov}(X,Y) \ .
\label{eqn:covsum}
\end{equation} 
Another general result for constants $a, b, c, d$ and random variables $W, X, Y, Z$ is
\begin{align}
& {\rm Cov}(aX+bY,cW+dZ) = \\ \nonumber & ac\,{\rm Cov}(X,W) + ad\,{\rm Cov}(X,Z) + bc\,{\rm Cov}(Y,W) + bd\,{\rm Cov}(Y,Z) \ .
\end{align} 
From this it follows that the covariance of two colours involving a common band, e.g. $m_1 - m_2$ and $m_2 - m_3$, assuming that each band is measured independently, is
\begin{equation}
{\rm Cov}(m_1-m_2, m_2 - m_3) = -{\rm Var}(m_2) \ .
\end{equation} 
The colour vector formed from the five bands $BVJHK$,
\begin{equation}
\vecp = (B-V, V-J, J-H, H-K) \ ,
\end{equation}
therefore has covariance matrix
\begin{equation}
{\mathbfss C}_p = 
\begin{pmatrix}
\sigma^2_B + \sigma^2_V  &   -\sigma^2_V              &   0              & 0       \\
-\sigma^2_V              &  \sigma^2_V + \sigma^2_J   &  -\sigma^2_J     &  0        \\
0                       &  -\sigma^2_J               &  \sigma^2_J + \sigma^2_H   &  -\sigma^2_H    \\
0                       &  0                         &  -\sigma^2_H     &  \sigma^2_H + \sigma^2_K   \\
\end{pmatrix}
\end{equation}
where $\sigma^2_i$ is the variance in band $i$.


\end{document}